\documentclass[12pt]{article}

\usepackage[margin=1.9cm]{geometry}

\usepackage{times}
\usepackage{graphicx}
\usepackage{amssymb}
\usepackage{amsthm}
\usepackage{amsmath}
\usepackage{dsfont}
\usepackage{bm}
\usepackage{mathrsfs}
\usepackage{bbold}

\usepackage[sorting=none,giveninits=true,maxcitenames=2,maxbibnames=2]{biblatex}
\AtEveryBibitem{
 \clearlist{language}
 \clearfield{issn}
 \clearfield{isbn}
 \clearfield{note}
}

\usepackage[labelfont=bf]{caption}
\usepackage[export]{adjustbox}

\usepackage[dvipsnames]{xcolor}
\usepackage{hyperref}
\hypersetup{
    colorlinks=true,
    linkcolor=Mahogany,
    citecolor=Blue,      
    urlcolor=cyan,
    linktoc=page}
\newcommand{\lcdm}{$\Lambda$CDM}

\newcommand{\EqRef}[1]{Eq.~({#1})}
\newcommand{\EqsRef}[1]{Eqs.~({#1})}

\newcommand{\formatdoi}[1]{\href{http://dx.doi.org/#1}{#1}}

\DeclareMathOperator{\al}{\alpha_{\Lambda}}

\begin{document}

\title{Lifshitz cosmology: quantum vacuum and Hubble tension}

\author{Dror Berechya\thanks{E-mail: \href{mailto:dror.berechya@weizmann.ac.il}{dror.berechya@weizmann.ac.il}}{ } and
Ulf Leonhardt\thanks{E-mail: \href{mailto:ulf.leonhardt@weizmann.ac.il}{ulf.leonhardt@weizmann.ac.il}}
\\
Weizmann Institute of Science, Rehovot 7610001, Israel}

\date{\today}

\maketitle

\begin{abstract}

Dark energy is one of the greatest scientific mysteries of today. The idea that dark energy originates from quantum vacuum fluctuations has circulated since the late '60s, but theoretical estimations of vacuum energy have disagreed with the measured value by many orders of magnitude, until recently. Lifshitz theory applied to cosmology has produced the correct order of magnitude for dark energy. Furthermore, the theory is based on well--established and experimentally well--tested grounds in atomic, molecular and optical physics. In this paper, we confront Lifshitz cosmology with astronomical data. We find that the dark--energy dynamics predicted by the theory is able to resolve the Hubble tension, the discrepancy between the observed and predicted Hubble constant within the standard cosmological model. The theory is consistent with supernovae data, Baryon Acoustic Oscillations and the Cosmic Microwave Background. Our findings indicate that Lifshitz cosmology is a serious candidate for explaining dark energy. 

\end{abstract}


\section{Introduction} \label{Introduction}

The cosmological standard model, the $\Lambda$ Cold Dark Matter (\lcdm{}) model, has been spectacularly successful. With a few basic principles, it explains a vast range of phenomena over an enormous range of time scales. With only six free parameters, it fits the complex and detailed fluctuation spectra of the cosmic microwave background (CMB). Nevertheless, the \lcdm{} model lacks an explanation of the underlying nature of three of its pillars, known as the dark sector --- inflation, dark matter, and dark energy.

In recent years, the cosmology community has been actively looking for cracks in the  \lcdm{} model in the form of tensions between several independent phenomena \cite{verde_tensions_2019}. Presently, the most severe such tension is known as the Hubble tension: the discrepancy between the Hubble constant (the present--day expansion rate) inferred from early--universe phenomena and the value obtained by local probes of cosmic expansion \cite{riess_expansion_2020,verde_tensions_2019}. Not everyone agrees that these tensions are real \cite{efstathiou_lockdown_2020} but by revealing cracks in the \lcdm{} model they may shed light on the dark sector.

There have been numerous attempts to explain the Hubble tension \cite{di_valentino_realm_2021}. Without exception, they either require significant changes to general relativity, the cosmological principle, or modifications to the standard model of particle physics that have not been experimentally tested elsewhere.

Here enters the Lifshitz theory in cosmology \cite{leonhardt_lifshitz_2019}. This theory is based on solid foundations in atomic, molecular, and optical (AMO) physics that have been experimentally tested with percent--level precision  \cite{decca_measuring_2014}. The connection to cosmology is the analogy between curved space--times and dielectric media \cite{plebanski_electromagnetic_1960,leonhardt_essential_2010} which is also the foundation of the well--developed field of transformation optics \cite{service_strange_2010}. A homogeneous and isotropic, expanding universe with scale factor $a(t)$ is perceived by the electromagnetic field as a medium with a homogeneous and isotropic but evolving refractive index $n(t) \propto a(t)$. Then, calculating the vacuum energy in the universe should be done as if it were a dielectric medium with an evolving refractive index in what is known as Lifshitz theory \cite{landau_statistical_1980,lifshitz_theory_1954}. Applied to cosmology, the Lifshitz vacuum energy turns out to have the same order of magnitude as the measured cosmological constant $\Lambda$ \cite{leonhardt_lifshitz_2019}.

In this paper, we compare the predictions of Lifshitz theory with astronomical data. We also formulate the theory such that that it can be taken up by astronomers. Lifshitz theory in cosmology has not been designed to alleviate the Hubble tension, but we show that the most naive choice of its coupling parameter fits the SH0ES value \cite{riess_cosmic_2021} with perfect precision. We also find that the theory is consistent with the Pantheon type Ia supernova (SN Ia) data at the same level or slightly better than the \lcdm{} model, that it agrees with the measured Baryon Acoustic Oscillations (BAO) and does not lead to deviations from the measured CMB spectra within the accuracy of the cosmic parameters. There are still many opportunities for further analysis, but the findings reported here already show that Lifshitz cosmology is a serious contender for a realistic explanation of dark energy, rooted in established physics.


\section{Lifshitz theory in cosmology} \label{The_Lifshitz_theory_in_cosmology}

\subsection{Background} \label{Background}

Most of our universe is empty space. Yet, this `emptiness' is far from being `nothingness.' According to the modern view of quantum field theory (QFT), the universe is filled with quantum fields in at least their ground state --- also known as the vacuum state. Since the early days of QFT, it is known that the vacuum state of a quantum field contains non--vanishing energy density, and due to Casimir in the late '40s, we know that this energy density may even exert measurable forces \cite{casimir_attraction_1948,casimir_influence_1948}. The physics of the quantum vacuum has been well--tested \cite{decca_measuring_2014,munday_measured_2009,rodriguez_casimir_2011,zhao_stable_2019} and explains a vast set of phenomena, from the adhesion of geckos to walls \cite{autumn_gecko_2008} to the limit trees can grow \cite{koch_limits_2004}.

So, the state of affairs is as follows. We know the universe is filled with quantum fields at their ground state, we know that this ground state exhibits non--vanishing energy density and may exert forces, and finally, we know that the universe is also filled with a mysterious energy density we call dark energy. It is therefore tempting to combine the physics of the quantum vacuum and dark energy.

Zel'dovich was the first to suggest, in 1968, that the cosmological constant $\Lambda$ comes from the physics of the quantum vacuum \cite{zeldovich_cosmological_1968}. By calculating the bare energy density of the vacuum, with a cut--off at the Planck scale where presumably GR breaks, one gets the correct structure of the cosmological constant. So, have we found an explanation of dark energy? Not quite yet. The problem is that the quantitative prediction of the vacuum energy density is off by about 120 order of magnitude \cite{weinberg_cosmological_1989}. Furthermore, if the theory is made to agree with the observed value of the vacuum energy density by choosing a sufficiently low cut--off for the vacuum fluctuations, it severely disagrees with measurements of vacuum forces \cite{mahajan_casimir_2006}.

This situation does not seem very encouraging. However, the case for a Casimir cosmology is not closed yet \cite{leonhardt_lifshitz_2019,leonhardt_case_2020}; the idea that dark energy stems from vacuum fluctuations \cite{zeldovich_cosmological_1968,weinberg_cosmological_1989,sakharov_vacuum_1991} may still be valid. The one encouraging insight is that curved space--times are the same as dielectric media in the eyes of the electromagnetic field: Maxwell's equations in curved space--time are equivalent to Maxwell's equations in dielectric media \cite{plebanski_electromagnetic_1960,leonhardt_essential_2010}, and our spatially--flat, expanding universe is just another curved space--time. It would be a far more unreasonable assumption that the universe is one particular space--time with different rules or that vacuum physics is different in the lab and the universe. Therefore, we assume that we can calculate vacuum energy in the universe as if it were the corresponding dielectric medium.

Now, since Zel'dovich, substantial progress has been made in understanding the quantum vacuum forces such that formal arguments can be replaced by empirically tested theory \cite{simpson_forces_2015,rodriguez_casimir_2011,scheel_casimir_2014}. Without exception, the empirical evidence for forces of the quantum vacuum and the comparison with theory comes from AMO physics. There the quantum vacuum produces attractive or repulsive forces \cite{munday_measured_2009,zhao_stable_2019} between dielectric objects and inside inhomogeneous media. For example, in the Casimir effect \cite{casimir_attraction_1948}, vacuum fluctuations cause two dielectric plates to attract each other. Here the spatial variation of the refractive index from free space to the material of the plates generates a vacuum force on the surface of each plate. This effect is a general phenomenon: variations of the refractive index create variations in the electromagnetic energy density and stress $\sigma$ in media \cite{scheel_casimir_2014,landau_statistical_1980,lifshitz_theory_1954,dzyaloshinskii_general_1961}, which gives the force density $\nabla\cdot\sigma$. This fact means that Casimir forces do not only act between dielectric bodies such as mirrors but also inside inhomogeneous bodies. Inhomogeneous dielectric media do exert local vacuum forces \cite{landau_electrodynamics_1995,griniasty_casimir_2017}.

The theory that agrees with modern measurements \cite{decca_measuring_2014} of Casimir forces is the Lifshitz theory \cite{rodriguez_casimir_2011,scheel_casimir_2014,landau_statistical_1980,lifshitz_theory_1954,dzyaloshinskii_general_1961}. Due to the analogy mentioned above between space--times and media, Lifshitz theory can be applied to cosmology; in that case, the electromagnetic field and its fluctuations perceive the (spatially--flat, homogeneous, and isotropic) expanding universe as a spatially--uniform but time--dependent medium with a refractive index that is proportional to the scale factor $a$ \cite{leonhardt_lifshitz_2019,leonhardt_case_2020}. Admittedly, when applying Lifshitz theory to that specific kind of medium, we extrapolate the theory outside its well--tested zone and introduce some new ideas. Nevertheless, the application of Lifshitz theory to the expanding universe was shown to produce the correct order of magnitude for the dark energy density \cite{leonhardt_lifshitz_2019}. 

In time--dependent media, the vacuum energy turns out to be time--dependent and responding to the evolution of the refractive index, or in the case of cosmology --- to the evolution of the universe. Then, by the Friedmann equations, the universe is reacting to the vacuum energy. In the following, we present the resulting self--consistent dynamics.

\subsection{Equations of motion} \label{Equations_of_motion}

In the framework of the flat--\lcdm{} model, the background (homogeneous and isotropic) universe evolves by the Friedmann equations, which can be summarized into one equation as
\begin{equation}
    H^2(a) = H_0^2(\Omega_r a^{-4} + \Omega_m a^{-3} + \Omega_{\Lambda}) \qquad \text{($\Lambda$CDM)}
    \label{LCDM_background_dynamics}
\end{equation}
where $H(a)$ is the Hubble parameter, $H_0$ the Hubble constant (Hubble parameter at the present--day), $a$ is the scale factor, and $\Omega_x$ with $x = r,m,\Lambda$ are the density parameters for radiation, matter, and the cosmological constant $\Lambda$.

Let us now see how this equation changes for the Lifshitz theory in cosmology (hereafter, `Lifshitz cosmology,' LC). For a given cosmic expansion, i.e., for a given $a(t)$, Lifshitz theory predicts for a medium with $n(t) \propto a(t)$ the energy--momentum tensor of the quantum vacuum in that medium \cite{leonhardt_lifshitz_2019}, in our case, in the universe. In turn, the vacuum energy and stress react back on the cosmic evolution through the Friedmann equation, influencing $a(t)$. This mutual interaction between the vacuum energy and the background universe results in self--consistent dynamics \cite{leonhardt_lifshitz_2019}, which we express here as\footnote{The dynamics that would result from the original calculations in Ref.~\cite{leonhardt_lifshitz_2019} are somewhat different from the dynamics we bring here; the reason for this difference is a different definition of the vacuum state. See Appendix~\ref{The_vacuum_state} for more details.}
\begin{equation}
    \Bigg\{
    \begin{aligned}
        H^2(a) \;\! \ &= H_0^2(\Omega_r a^{-4} + \Omega_m a^{-3} + \Omega_{LC}) \\
        H_0^2 \Dot{\Omega}_{LC} &= 8\alpha_\Lambda H \partial_t^3 H^{-1} \qquad \qquad \qquad \, \text{(Lifshitz cosmology)}
    \end{aligned}
    \label{LC_self_consistent_dynamics}
\end{equation}
where $\Omega_{LC}$ is the density parameter for dark energy in the Lifshitz cosmology. $\alpha_\Lambda$ is a dimensionless coupling parameter that depends on the cut--off, assumed near the Planck scale, and on the possible contributions of other fields in the standard model of particle physics \cite{leonhardt_lifshitz_2019}. As these influences are not known within the present theory, $\alpha_\Lambda$ is a free parameter that must be fitted against observations. Taking only the electromagnetic field into account and assuming a sharp cut--off at exactly the Planck scale, we get \cite{leonhardt_lifshitz_2019} $\alpha_\Lambda^{TH}=(9\pi)^{-1}$ (the `\textsc{th}' superscript indicates that this value is a theoretical prediction under the above--mentioned conditions). A dot above a character denotes differentiation with respect to cosmological time $t$.

The second equation in \EqsRef{\ref{LC_self_consistent_dynamics}} describes the response of the vacuum energy to changes of the scale factor $a$ (or, in the language of the Lifshitz theory, the refractive index). This equation hides an integration constant, which remains a free parameter that must also be fitted against observations. Thus, Lifshitz cosmology replaces one of the \lcdm{} parameters, namely $\Omega_\Lambda$, with two new parameters: $\al$ and the integration constant (giving us a total of only seven parameters).


\section{Approximate solution} \label{Approximate_Solution}

In Sec.~\ref{Background}, we have presented the ideas behind the description of dark energy as the vacuum energy produced in a time--dependent dielectric medium using Lifshitz theory (for further details, see Refs.~\cite{leonhardt_lifshitz_2019} and \cite{leonhardt_case_2020}). In Sec.~\ref{Equations_of_motion}, we saw that this theory also gives a testable prediction: a modified expansion history, embodied in \EqsRef{\ref{LC_self_consistent_dynamics}}. In this section, we will find an approximate solution for the dynamics predicted by Lifshitz cosmology. Later, we will use this approximate solution to analyze the dynamics and demonstrate the theory's plausibility. Here we present the approximate solution along general lines; for further details, see Appendix~\ref{calculations}.

The contribution of the cosmological constant in the \lcdm{} model is negligible at last--scattering as $\Omega_\Lambda/\Omega_m (1+z_*)^3 \approx 1.7\cdot10^{-9}$ with values provided by the \emph{Planck collaboration}\footnote{Throughout the paper, we will use italic letters to designate the \href{https://www.cosmos.esa.int/web/planck}{\emph{Planck collaboration}} and distinguish it from Planck the person or other contexts in which this name might appear.} \cite{planck_collaboration_planck_2020}, and it is even smaller before that time. We assume that in Lifshitz cosmology the vacuum contribution is negligible before last--scattering as well and verify this later. The right--hand side of the second equation in \EqsRef{\ref{LC_self_consistent_dynamics}} is zero for linear $H^{-1}$; this means that $\Omega_{LC}$ is constant in both radiation-- and matter--domination eras where $H^{-1}$ is linear. Lifshitz cosmology may intervene only in the transition period around $a_{eq}$ (as we will see in detail in Sec.~\ref{Implications_for_the_early_universe}). Hence, if we start with a negligible vacuum contribution during radiation domination, then the vacuum contribution will remain negligible at the beginning of matter--domination if the effects of Lifshitz cosmology around $a_{eq}$ are small. Here, we assume that this is the case; in Sec.~\ref{Implications_for_the_early_universe} we check the validity of this assumption (see Fig.~\ref{RelContZ}). Thus, we adopt \lcdm{}'s dynamics at the early universe and focus our attention on the late universe. Hence we drop the radiation term in our calculations (as it turns out, see Appendix~\ref{calculations}, this is a crucial simplification for our calculations).

Even without the radiation term, finding a closed analytical solution for \EqsRef{\ref{LC_self_consistent_dynamics}} remains a real challenge. Moreover, finding a numerical solution is no less challenging, mainly for the following two reasons. First, \EqsRef{\ref{LC_self_consistent_dynamics}} are ``stiff equations,'' causing havoc with step size and accuracy, and second, the equation for $\Dot{\Omega}_{LC}$ depends on high derivatives of $a$ (up to fourth--order), which is problematic since the highest derivatives take the lead in differential equation solvers. In reality, the dynamics of $\Omega_{LC}$ are a mere correction to the dynamics of the universe.

Therefore, we solve for the dynamics after last--scattering \emph{perturbatively}. $\al$ is presumably small (recall that the theoretical prediction is $\alpha_\Lambda^{TH}=(9\pi)^{-1} \approx 0.035$), so we calculate $\Omega_{LC}$ up to first--order in $\al$. We plug the zeroth--order Hubble parameter (the \lcdm{}'s Hubble parameter, \EqRef{\ref{LCDM_background_dynamics}} without the radiation term) into the equation for $\Dot{\Omega}_{LC}$ (the second equation in \EqsRef{\ref{LC_self_consistent_dynamics}}), and we integrate it (analytically, see Appendix~\ref{calculations}) with $\Omega_{\infty} \equiv \displaystyle \lim_{a \rightarrow \infty}\Omega_{LC}$ as the integration constant. In this way, we get the first--order correction to $\Omega_{LC}$, that we substitute into the equation for $H(a)$ [the first equation in \EqsRef{\ref{LC_self_consistent_dynamics}}]. Thus, we get
\begin{align}
    \nonumber H^2 &= H_0^2(\Omega_m a^{-3} + \Omega_{LC}), \\
    \nonumber \Omega_{LC} &= \Omega_{\infty} \left[1+18\al\left(\ln{(\frac{\Omega_m}{\Omega_{\infty}}a^{-3}+1)}-\frac{3}{\frac{\Omega_{\infty}}{\Omega_m}a^{3}+1}\right)\right] \\
    &\qquad \qquad \qquad \qquad \qquad \qquad \qquad \qquad \quad \text{(for $a_{eq} \ll a$)}.
    \label{First_order_H_after_equality}
\end{align}

\begin{figure}
    \centering
    \includegraphics{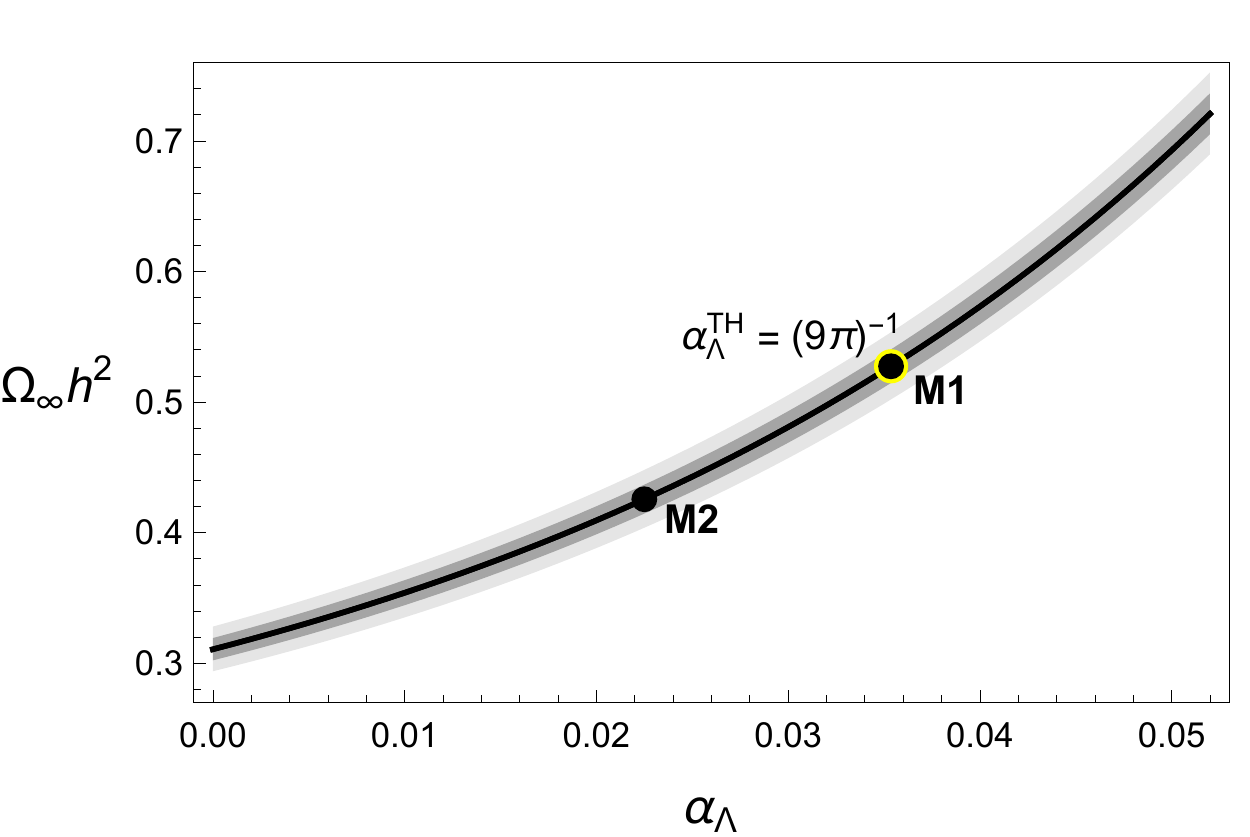}
    \caption{$\Omega_{\infty} h^2 \equiv \displaystyle \lim_{a \rightarrow \infty}\Omega_{LC}h^2$ as a function of $\al$ in the range of interest. This relationship results when demanding that Lifshitz cosmology preserves $D_M$ (\EqRef{\ref{fixing_the_integration_constant}}), and it is calculated according to \EqRef{\ref{fitting_condition_for_delta}} with $\omega_m^P = 0.1430$ and $\omega_\Lambda^P = 0.3107$ from Ref.~\cite{planck_collaboration_planck_2020}. The bands show the $\pm 1 \sigma$ (dark gray) and $\pm 2 \sigma$ (light gray) errors in $\Omega_{\infty} h^2$ as estimated by propagating the errors in $D_M^{(\Lambda \mathrm{CDM})}$ (see Appendix~\ref{Late_universe}). The two realizations of Lifshitz cosmology considered in this paper are also shown: M1 (black and yellow point), the theoretical prediction for electromagnetic contribution alone with a sharp cut--off at exactly the Planck length, $\alpha_\Lambda^{M1} = \alpha_\Lambda^{TH} = (9\pi)^{-1}$ and M2 (black point), $\alpha_\Lambda^{M2} = 0.0225$.}
    \label{OmegaInfVSalpha}
\end{figure}

The next step is to fit the theory's parameters with cosmological data sets, such as CMB power spectra, SN Ia, and BAO. The complete way of fitting the parameters is to include the modified equation for the background dynamics, i.e., the new equation for $H(a)$, in the relevant computer codes and perform statistical analysis (such as likelihood--based MCMC or Fisher information). In this paper, however, we aim to explore the Lifshitz dynamics for the first time to test whether this theory can plausibly resolve the Hubble tension at all, which would then justify further research. For this, we take the \lcdm{}'s value for the sound horizon, $r_*$, as we assume that the deviations of Lifshitz cosmology from the \lcdm{} model are negligible in the early universe. The following section shows that the resulting dynamics are consistent with BAO and SN Ia measurements. To preserve the acoustic angular scale of the CMB fluctuations ($\theta_* \equiv r_* / D_M$), we must demand that the angular diameter distance to the surface of last--scattering, $D_M$, is unchanged as well:
\begin{equation}
    D_M = c \int^{z_*}_0 \frac{dz'}{H(z')} = D_M^{(\Lambda \mathrm{CDM})}
    \label{fixing_the_integration_constant}
\end{equation}
where $D_M^{(\Lambda \mathrm{CDM})}$ is calculated with the \lcdm{} model. In effect, this demand gives us a relationship between $\al$ and the combination $H_0^2\Omega_{\infty}$ for the following reason. Since $H_0^2\Omega_m$ is proportional to the physical matter density, it should be a model--independent quantity; therefore, we may use \lcdm{}'s value for this combination. The \emph{Planck collaboration} determined $\omega_m^P \equiv [\Omega_m h^2]^P = 0.1430 \pm 0.0011$ \cite{planck_collaboration_planck_2020} via the relative heights of the CMB acoustic peaks (approximately) model--independently \cite{planck_collaboration_planck_2014}. Here and throughout this paper $h \equiv H_0/100 [\mathrm{km \: s^{-1} \: \! Mpc^{-1}}]$ and the `\textsc{p}' superscript, hereafter, denotes that the value is determined by \emph{Planck}. Figure~\ref{OmegaInfVSalpha} shows the resulting relationship. The $\Omega_{\infty} h^2$ errors presented in this figure are estimates based solely on propagating the \lcdm{} errors in determining $D_M^{(\Lambda \mathrm{CDM})}$; that is, for any other \lcdm{}'s quantity, we take the mean value given by \emph{Planck}'s TT,TE,EE+LowE+lensing analysis \cite{planck_collaboration_planck_2020} without errors (see Appendix~\ref{Late_universe} for more details).

\begin{figure}
    \centering
    \includegraphics{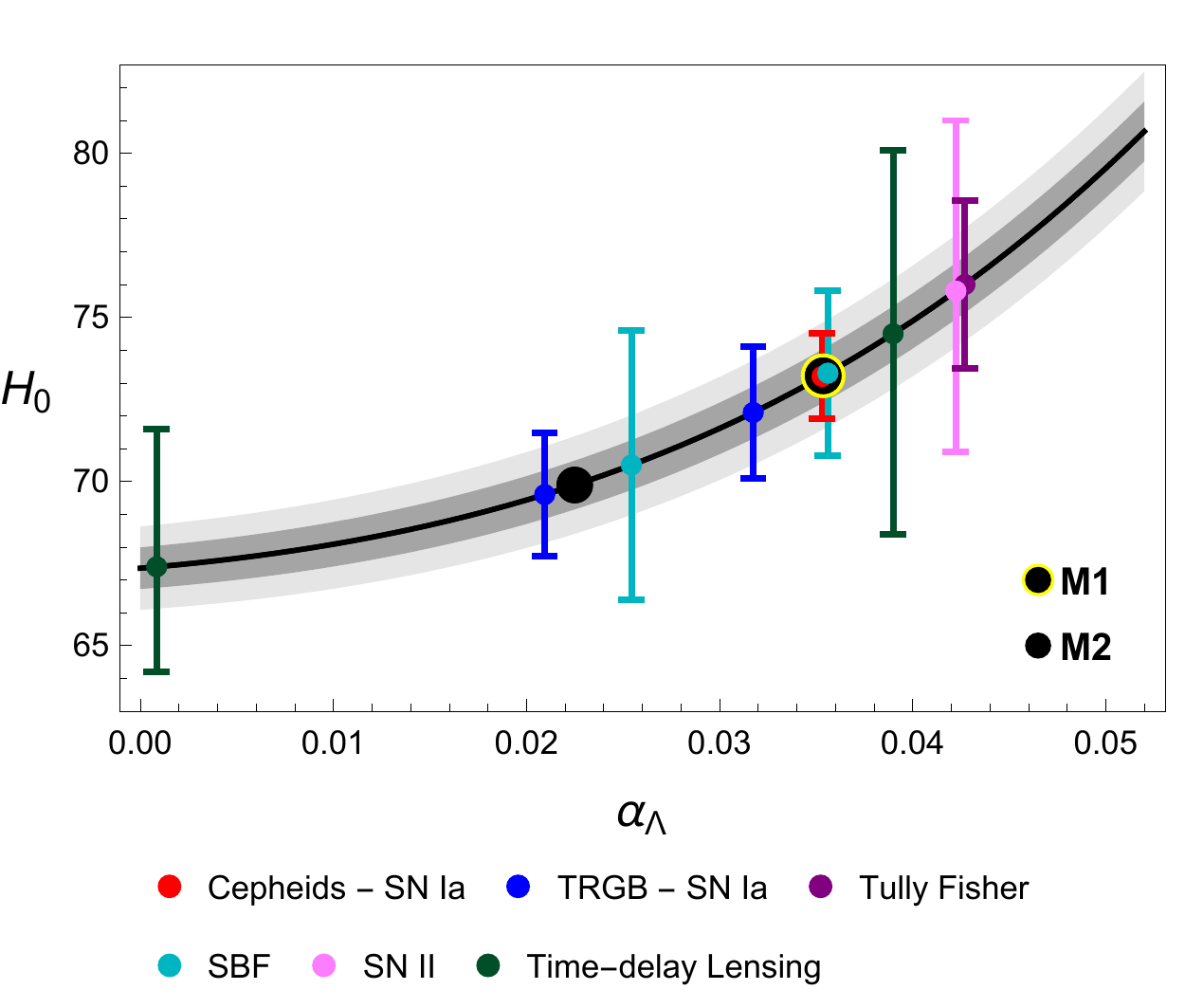}
    \caption{$H_0$ in units of $\mathrm{km \: s^{-1} \: \! Mpc^{-1}}$ as a function of $\al$ in the range of interest.     
    The bands show the $\pm 1 \sigma$ (dark gray) and $\pm 2 \sigma$ (light gray) errors in $H_0$ obtained by propagating the errors in $\Omega_{\infty} h^2$. The theory points M1 and M2 are as in Fig.~\ref{OmegaInfVSalpha}. We also show several local measurements of $H_0$, done by several independent groups using several independent methods: Cepheids - SN Ia ($73.2\pm1.3$ by \citeauthor{riess_cosmic_2021}, \citeyear{riess_cosmic_2021} \cite{riess_cosmic_2021}, SH0ES team), TRGB - SN Ia ($72.1\pm2.0$ by \citeauthor{soltis_parallax_2021}, \citeyear{soltis_parallax_2021} \cite{soltis_parallax_2021} and $69.6\pm1.88$ by \citeauthor{freedman_calibration_2020}, \citeyear{freedman_calibration_2020} \cite{freedman_calibration_2020}), Tully Fisher ($76.00\pm2.55$ by \citeauthor{kourkchi_cosmicflows-4_2020}, \citeyear{kourkchi_cosmicflows-4_2020} \cite{kourkchi_cosmicflows-4_2020}), Surface Brightness Fluctuations (SBF) ($73.3\pm2.5$ by \citeauthor{blakeslee_hubble_2021}, \citeyear{blakeslee_hubble_2021} \cite{blakeslee_hubble_2021} and $70.5\pm4.1$ by \citeauthor{khetan_new_2021}, \citeyear{khetan_new_2021} \cite{khetan_new_2021}), SN II ($75.8^{+5.2}_{-4.9}$ by \citeauthor{dejaeger_measurement_2020}, \citeyear{dejaeger_measurement_2020} \cite{dejaeger_measurement_2020}), and Time-delay Lensing ($74.5^{+5.6}_{-6.1}$ and $67.4^{+4.2}_{-3.2}$ by \citeauthor{birrer_tdcosmo_2020}, \citeyear{birrer_tdcosmo_2020} \cite{birrer_tdcosmo_2020}). All the values are in units of $\mathrm{km \: s^{-1} \: \! Mpc^{-1}}$ and quoted from the compilation in Ref.~\cite{divalentino_combined_2021}. As can be seen, whatever the actual value of $H_0$ is, Lifshitz cosmology may reproduce the correct value (at least nominally, see the discussion in Sec.~\ref{The_Hubble_diagram_and_distance_ladders}). The theoretical prediction (M1) is more or less at the middle of the local measurements, and remarkably, it is right on the latest measurement by the SH0ES team (red point).}
    \label{H0VSalpha}
\end{figure}

Thus, we are left only with $\al$ as a free parameter. The value of $\al$ will determine $H(a)$, and hence, will fix the value of $H_0$, Fig.~\ref{H0VSalpha}, as well as the values of the other parameters in Table~\ref{parameters}.

Figure~\ref{H0VSalpha} shows that whatever the actual value of $H_0$ may be, Lifshitz cosmology may reproduce it (at least nominally, see the discussion in Sec.~\ref{The_Hubble_diagram_and_distance_ladders}). The theoretical prediction under the assumptions of only electromagnetic contributions and a sharp cut-off at exactly the Planck length is more or less at the middle of the local measurements, and remarkably, it is right on the latest measurement by the SH0ES team \cite{riess_cosmic_2021}.

To study the influence of different values of $\al$ and hence of different sets of parameters, we will explore the resulting dynamics of two representative realizations of Lifshitz cosmology. The first one, which we call `M1,' is the theoretical prediction, for which we have $\alpha_\Lambda^{M1} = \alpha_\Lambda^{TH} = (9\pi)^{-1}$. For the second realization, which we call `M2,' we take $\alpha_\Lambda^{M2} = 0.0225$. Here we choose two values for $\al$ as examples, and then $\Omega_{\infty} h^2$ is determined by \EqRef{\ref{fixing_the_integration_constant}}. The obtained parameter values for the two realizations M1 and M2 are presented in Table~\ref{parameters}. We regard these two realizations as examples for the viability of the theory.

\begin{table}
    \centering
    \caption{Two realizations of Lifshitz cosmology. For each realization, we choose $\al$. Then $\Omega_{\infty} h^2$ is fixed by \EqRef{\ref{fixing_the_integration_constant}} (taking $\omega_m^P = 0.1430$ and $\omega_\Lambda^P = 0.3107$ \cite{planck_collaboration_planck_2020}, see Appendix~\ref{calculations} for details of the calculations), and the rest of the parameters of Lifshitz cosmology follow. The errors in the parameters result from propagating the errors in $\Omega_{\infty} h^2$ that are estimated by propagating the errors in $D_M^{(\Lambda \mathrm{CDM})}$ while calculating $\Omega_{\infty} h^2$.}
    \begin{adjustbox}{center}
    \begin{tabular}{c | c c c c c c c}
        \hline
        \hline
         & $\al$ & $H_0 \left[\frac{\mathrm{km \: s^{-1}}}{\mathrm{Mpc}}\right]$ & $\Omega_{\infty} h^2$ & $\Omega_{LC}(z=0) h^2$ & $\Omega_m h^2$ & $\Omega_{LC}(z=0)$ & $\Omega_m$ \\ \hline
        M1 & $(9\pi)^{-1}$ & $73.2 \pm 0.8$ & $0.527 \pm 0.013$ & $0.393 \pm 0.012$ & $0.143 \pm 0.017$ & $0.733 \pm 0.028$ & $0.267 \pm 0.028$ \\
        M2 & $0.0225$ & $69.9 \pm 0.7$ & $0.426 \pm 0.011$ & $0.345 \pm 0.010$ & $0.143 \pm 0.015$ & $0.707 \pm 0.026$ & $0.293 \pm 0.026$ \\
        \hline
    \end{tabular}
    \end{adjustbox}
    \label{parameters}
\end{table}


\section{Low redshift probes} \label{BAO_SN_Ia_and_distance_ladders}

At this point, we have in our hands a theory explaining the physical origin of dark energy, one that stems from well--known physics, an approximate solution for the theory's dynamics assuming unmodified early evolution, and two sets of parameters (in Table~\ref{parameters}) for two realizations of the theory: M1 ($\alpha_\Lambda^{M1} = (9\pi)^{-1}$, a theoretical prediction) and M2 ($\alpha_\Lambda^{M2} = 0.0225$).

Now, we are ready to compare the resulting dynamics with low redshift probes of cosmic expansion, viz. SNe Ia and BAOs. We shall see that M1 fits better the SN data with SH0ES calibration of the absolute magnitude $M_B$, which might be crucial regarding the Hubble tension \cite{efstathiou_h0_2021,benevento_can_2020}; however, M1's fit to BAO data is somewhat lesser than \lcdm{}'s. On the other hand, M2 fits BAO data better than M1 and seemingly falls from \lcdm{}'s fit only by a small margin; yet, M2 fits SN data with a lower value (more negative) of $M_B$ and can only relieve the tension (see the discussion in the following subsection).

We will compare the resulted dynamics of M1 and M2 with observational findings, refraining from a more complex statistical analysis for the time being. Our analysis already indicates the viability of Lifshitz cosmology. However, only a complete statistical analysis will determine the actual set of values for the theory's parameters, instead of M1 and M2, which are demonstrations obtained by choosing $\al$ and imposing \EqRef{\ref{fixing_the_integration_constant}}, and will enable us to decide which is the better theory. This further analysis poses an opportunity for future research.

\subsection{The Hubble diagram and distance ladders} \label{The_Hubble_diagram_and_distance_ladders}

We start with SNe Ia observations. Ultimately, each SN Ia measures the luminosity distance via the relation
\begin{equation}
    \mu \equiv m_B - M_B + \delta\mu = 5\log_{10}\frac{D_L(z)}{\mathrm{Mpc}} + 25,
    \label{distance_modulus_definition}
\end{equation}
where
\begin{equation}
    D_L(z) = c \, (1+z) \int_0^z \frac{dz'}{H(z')} \,,
    \label{luminosity_distance}
\end{equation}
$\mu$ is the distance modulus, $M_B$ is the absolute magnitude (in the B--band), $m_B$ is the apparent magnitude, and $\delta\mu$ summarizes corrections due to effects such as color, light--curve's shape, and host--galaxy mass; these effects can be either measured or fitted using SN Ia data alone, independently of cosmology \cite{riess_type_2018}. Roughly speaking, in each measurement, we measure $m_B$ and $z$, and we wish to infer $D_L(z)$. $M_B$ is thus a nuisance parameter that must be determined or marginalized over. This nuisance parameter is degenerate with $H_0$ in the SN Ia data. As a prefactor in $H(z)$, $H_0$ would shift $M_B$ by $5\log_{10} H_0$ in \EqRef{\ref{distance_modulus_definition}}; thus, both $M_B$ and $H_0$ get swallowed into the intercept of the magnitude--redshift relation.

One way to break that degeneracy is to use a distance ladder to infer $M_B$ by calibrating SN Ia. Generally, there are two approaches to measuring $H_0$ using distance ladders. One is to use geometrical measurements to anchor local probes of distance (first rung), such as Cepheids (e.g., SH0ES team, \citeauthor{riess_cosmic_2021} \cite{riess_cosmic_2021}) or TRGB (e.g., \citeauthor{soltis_parallax_2021} \cite{soltis_parallax_2021} and \citeauthor{freedman_calibration_2020} \cite{freedman_calibration_2020}), then use these probes and go farther to calibrate the absolute magnitude of SN Ia in the same host galaxies (second rung), and finally, go farther still and use the calibrated absolute magnitudes to infer $H_0$ from SNe Ia in the Hubble flow (third rung). This approach is known as the \emph{local} distance ladder. The other approach starts instead with BAO standard rulers (assuming a value for $r_d$, see Sec.~\ref{Avoiding_the_sound_horizon_problem}, which makes this approach model--dependent) to calibrate the absolute magnitude of far--away supernovae; then, it uses the calibrated absolute magnitudes to infer $H_0$ from lower redshift SNe Ia. This approach is known as the \emph{inverse} distance ladder \cite{efstathiou_h0_2021}.

\begin{figure}[h]
    \centering
    \includegraphics[width=0.8\textwidth]{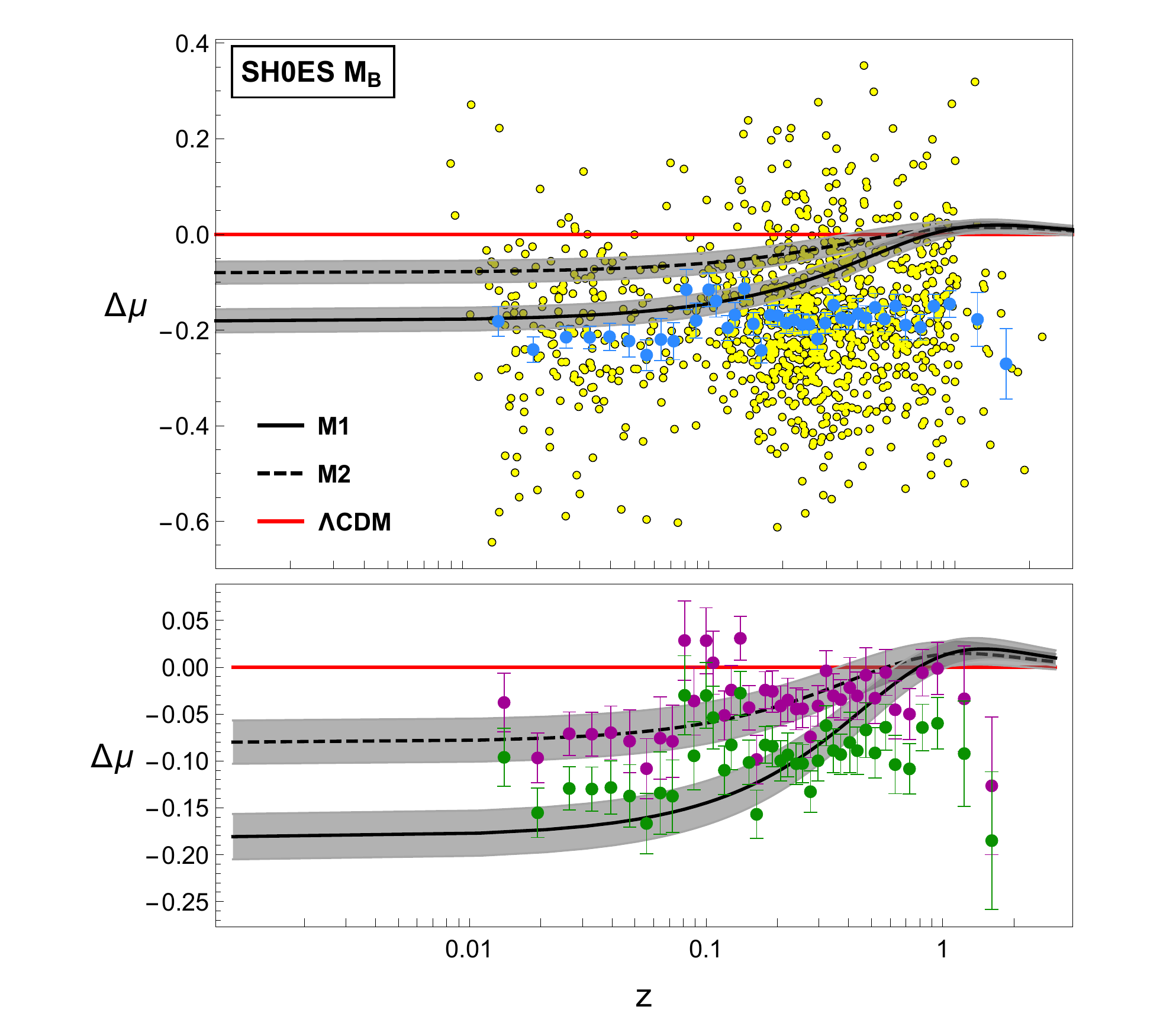}
    \caption{$\Delta\mu \equiv \mu - \mu_{\Lambda \mathrm{CDM}}$ as a function of $z$. Top panel: Pantheon data \cite{scolnic_complete_2018} (yellow (unbinned data) and blue (binned data) points) with the reproduced SH0ES absolute magnitude $M_B = -19.244$ from Ref.~\cite{efstathiou_h0_2021}. The theoretical predictions are also shown: unbroken black cure represents M1, and dashed black curve represents M2. The gray band around each curve shows the $\pm 1 \sigma$ errors in $\Delta\mu$ obtained by propagating the errors in $\Omega_{\infty} h^2$. Bottom panel: The binned Pantheon data are shown with $M_B = -19.330$ (Green) and with $M_B = -19.388$ (Purple), the best $M_B$ in terms of RMSD (\EqRef{\ref{RMSD_equation}}) for M1 and M2, respectively.}
    \label{muVSzFull}
\end{figure}

The SH0ES team takes the approach of the local distance ladder and uses Cepheids to calibrate the absolute magnitude of SN Ia. To reproduce SH0ES $M_B$, Ref.~\cite{efstathiou_h0_2021} combined the geometrical distance estimates of the maser galaxy NGC 4258 \cite{reid_improved_2019}, detached eclipsing binaries in the Large Magellanic Cloud \cite{pietrzynski_distance_2019}, and parallax measurements for 20 Milky Way Cepheids \cite{benedict_hubble_2007,van_leeuwen_cepheid_2007,riess_new_2018}, the SH0ES Cepheid photometry and Pantheon SN peak magnitudes, then Ref.~\cite{efstathiou_h0_2021} finds (Eq.~(6) there)
\begin{equation}
    M_B = -19.244 \pm 0.042 \mathrm{\: mag}.
    \label{absolute_magnitude_local}
\end{equation}
We have adopted this value and used it with the Pantheon data set (which is given by \citeauthor{scolnic_complete_2018}, \citeyear{scolnic_complete_2018} \cite{scolnic_complete_2018} and publicly available in doi:~\formatdoi{10.17909/T95Q4X}) to extract the $\mu$'s of observed SNe. We also calculated $\mu$ (\EqRef{\ref{distance_modulus_definition}}) for \lcdm{} and the two Lifshitz cosmologies (M1 and M2). Figure~\ref{muVSzFull} shows $\Delta\mu \equiv \mu - \mu_{\Lambda \mathrm{CDM}}$ for the theories and the observed data points with $M_B = -19.244$ (top panel). This figure also shows binned data from Ref.~\cite{scolnic_complete_2018}.

It has been noted \cite{efstathiou_h0_2021,benevento_can_2020} that, in principle, SH0ES does not measures $H_0$ directly but measures $M_B$ instead; $H_0$ is inferred from the low redshift ($z<0.15$ \cite{efstathiou_h0_2021,benevento_can_2020}) SNe in the Pantheon sample with the measured $M_B$. According to this view, the Hubble tension is really an $M_B$ tension: a significant gap of about $\Delta M_B \approx 0.2$ between the SH0ES $M_B$ and the one inferred from the Pantheon data without including the SH0ES constraint (retaining $M_B$ in the likelihood) \cite{benevento_can_2020} or the one obtained by inverse distance ladder \cite{efstathiou_h0_2021}. For theories that modify the dynamics above $z \approx 0.15$, these two viewpoints should be equivalent; however, for theories that modify the dynamics below that redshift, only the latter viewpoint ($M_B$ tension) should be considered, as in this case, $H_0$ is not constrained by the Pantheon data (see figure 1 in Ref.~\cite{benevento_can_2020}) and SH0ES analysis would be oblivious to this modification \cite{efstathiou_h0_2021,benevento_can_2020}. That is, if our universe would evolve according to a theory that modifies the dynamics below $z \approx 0.15$, it will not appear in the SH0ES analysis, and they would approximately measure the same \lcdm{} value for $H_0$ as inferred from the CMB.

Even though Lifshitz cosmology starts to modify the dynamics at $z>0.15$ (see Fig.~\ref{BAOdata}), we would like to estimate how the theory will perform regarding the $M_B$ tension. To give some quantitative measure, for each model, we calculate the root--mean--square deviation (RMSD) given by
\begin{equation}
    \mathrm{RMSD} = \sqrt{\frac{\sum^N_{i=1} (\mu(z_i)-\mu_i)^2}{N}}
    \label{RMSD_equation}
\end{equation}
where $z_i$ and $\mu_i$ are, respectively, the measured redshift and distance modulus (with a given $M_B$) of measurement $i$, $\mu(z_i)$ is the theoretical prediction, and $N$ is the number of data points (for the Pantheon sample, $N = 1,048$). For each model, we have found $M_B$ that gives the lowest RMSD using the (unbinned) Pantheon data. For comparison, we also calculated the RMSD with the reproduced SH0ES absolute magnitude ($M_B = -19.244$) \cite{efstathiou_h0_2021} and with absolute magnitude obtained by fitting a late dark energy model with \emph{Planck}, BAO, and Pantheon data (retaining $M_B$ in the Pantheon likelihood without the SH0ES constraint), $M_B = -19.415$ \cite{benevento_can_2020}. The results are presented in Table~\ref{RMSD_table}. The bottom panel of Fig.~\ref{muVSzFull} shows $\Delta\mu$ for the binned Pantheon data with the best $M_B$ of M1 and M2, together with the theoretical curves. We find $M_B^* = -19.421$ for \lcdm{}, $M_B^* = -19.330$ for M1, and $M_B^* = -19.388$ for M2 (the `$*$' superscript indicates a value corresponding to the lowest RMSD). The three RMSD values that correspond to the three $M_B^*$'s are comparable to one another, so it seems that the three models fit the unbinned Pantheon data comparably well.

These results show that while M2 can only mitigate the $M_B$ tension by about $19\%$ at mean value, as $\Delta M_B^* \equiv M_B^{(\mathrm{SH}0\mathrm{ES})} - M_B^* = 0.144$, M1 can relieve it considerably by about $51\%$ at mean value, as $\Delta M_B^* = 0.086$ (for \lcdm{}, one gets $\Delta M_B^* = 0.177$). On the other hand, M2 fits the shape of the binned Pantheon data exceedingly well, as shown in Fig.~\ref{muVSzFull}, while M1's fit to the shape is only moderate. The shape of the binned data might depend on the model, e.g., via model dependence of the redshift weights of the surveys \cite{benevento_can_2020}; in addition, the RMSD($M_B$) profile (for the unbinned data) is shallow around the minimum, so the two Lifshitz cosmologies might do even better (this seem to be especially true for M1). Only a rigorous statistical analysis would be able to tell. Nevertheless, based on our current analysis, we conclude that the two Lifshitz cosmologies fit the Pantheon sample as well as \lcdm{} does, and they both reduce the $M_B$ tension (M1 might even, hopefully, resolve this tension). As we will see later, while both cosmologies (M1 and M2) seem to fit the BAO data comparably to \lcdm{}, M2 is a better fit there.

\vspace{1.5cm}

\begin{table}
    \centering
    \caption{Root--mean--square deviation (RMSD) calculated with \EqRef{\ref{RMSD_equation}} for the (unbinned) Pantheon data \cite{scolnic_complete_2018}. The top block shows the best (in terms of RMSD) $M_B$ and the corresponding RMSD for each model. The bottom block shows results with the reproduced SH0ES absolute magnitude ($M_B = -19.244$) \cite{efstathiou_h0_2021} and the absolute magnitude obtained by fitting a late dark energy model with \emph{Planck}, BAO, and Pantheon data ($M_B = -19.415$) \cite{benevento_can_2020}.}
    \begin{tabular}{c | c c c}
        \hline
        \hline
        $M_B$ & \lcdm{} & M1 & M2 \\
        \hline
        $-19.421$ & 0.1449 & --- & --- \\
        $-19.330$ & --- & 0.1511 & --- \\
        $-19.388$ & --- & --- & 0.1453 \\
        \hline
        $-19.244$ & 0.2291 & 0.1736 & 0.2046 \\
        $-19.415$ & 0.1450 & 0.1736 & 0.1477 \\
        \hline
    \end{tabular}
    \label{RMSD_table}
\end{table}

\newpage

\subsection{Distance--ladder--independent analysis} \label{Ez_from_distance_ladder_independent_SNe_Ia_measurements}

Before we turn to BAO data, let us compare our dynamics with measurements of $E(z) \equiv H(z)/H_0$ that are independent of any distance ladder, going around the $M_B$ dilemma. The quantity $E(z)$ is independent of $H_0$ and thus avoids the $H_0$--$M_B$ degeneracy; therefore, it can be measured using SN Ia data alone \cite{riess_type_2018}.

\begin{figure}[t]
    \centering
    \includegraphics{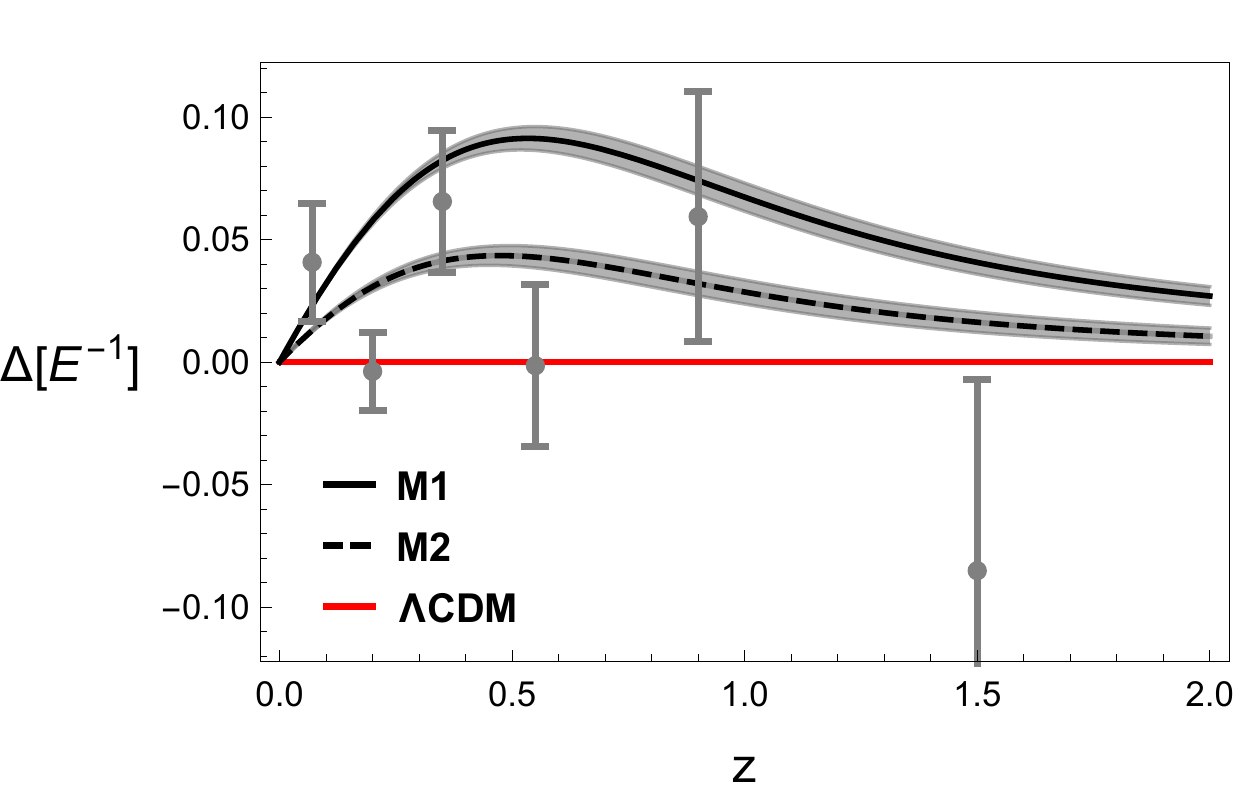}
    \caption{$\Delta[E^{-1}] \equiv E^{-1} - E^{-1}_{\Lambda \mathrm{CDM}}$ as a function of $z$. The six gray points are model--independent measurements of $E^{-1}$ performed by Ref.~\cite{riess_type_2018} based on SN Ia data alone. The theoretical predictions are also shown: unbroken black cure represents M1, and dashed black curve represents M2. The gray band around each curve shows the $\pm 1 \sigma$ errors in $\Delta[E^{-1}]$ obtained by propagating the errors in $\Omega_{\infty} h^2$. Among the six data points, three (at $z = 0.07, 0.35,$ and $0.9$) are situated closer to M1's curve, and the remaining three (at $z = 0.2, 0.55,$ and $1.5$) are situated closer to the \lcdm{} baseline. M2 lies in between, and it seems to agree with all data points. All in all, Lifshitz cosmology appears to fit the data comparably to \lcdm{}.}
    \label{Eofzdata}
\end{figure}

To extract $E(z)$ from the SN Ia data, Ref.~\cite{riess_type_2018} parametrize it by its value at several specific redshifts and interpolate to define the complete $E(z)$ function, which can then be used to compute the luminosity distance and compare to the data while fully marginalizing over the absolute magnitude. This way, they constrained the value of $E^{-1}(z)$ at six different redshifts model--independently (except for assuming a spatially--flat universe). Figure~\ref{Eofzdata} shows $\Delta[E^{-1}] \equiv E^{-1} - E^{-1}_{\Lambda \mathrm{CDM}}$ for these six data points together with the two Lifshitz cosmologies. It can be seen that among the six data points, three (at $z = 0.07, 0.35,$ and $0.9$) are situated closer to M1's curve, and the remaining three (at $z = 0.2, 0.55,$ and $1.5$) are closer to the \lcdm{}'s baseline. M2 lies in between, and it seems to agree with all data points. All in all, we conclude that Lifshitz cosmology does fit the $E^{-1}$ data to a degree comparable to \lcdm{}.

\subsection{BAO measurements} \label{Avoiding_the_sound_horizon_problem}

Now we turn to BAOs, the second low redshift probe we consider in this paper. BAO measurements can be used for measuring $H(z)$, as these measurements constrain the product $H(z) r_d$ \cite{di_valentino_realm_2021,arendse_low-redshift_2019}, where $r_d$ is the sound horizon at the end of the baryon--drag epoch ($z_d = 1,059.94$ \cite{planck_collaboration_planck_2020}).

The BAO constraint (at $z \ge 0.38$) on $H(z) r_d$ can be extrapolated to $z=0$ using a lower redshift probe, such as SN Ia, to obtain a constraint on $H_0 r_d$ \cite{arendse_low-redshift_2019}. This procedure of extrapolating the BAO measurements is model--dependent \cite{arendse_low-redshift_2019}. Nonetheless, the extrapolation can be performed using various cosmographic techniques, such as cosmology--agnostic expansions of the Hubble parameter or distances, so that the final measurement might be considered as independent of a cosmological model \cite{arendse_low-redshift_2019}. Therefore, a point has been made that due to the extrapolated BAO constraint on $H_0 r_d$, one cannot rise $H_0$ without reducing $r_d$ since this would introduce tension with the extrapolated BAO measurements \cite{arendse_low-redshift_2019}.

However, while the cosmographic techniques are agnostic to cosmology, they are still models, and it is not clear how well they may capture the Lifshitz cosmology. Ref.~\cite{di_valentino_realm_2021} has noted that the BAO data are extracted under the assumption of a \lcdm{} scenario, so one should be careful in excluding all the `Late Time solutions' only using this argument. Moreover, the use of SN Ia to extrapolate the BAO measurements in this inverse distance ladder procedure might be problematic, as also noted by Ref.~\cite{di_valentino_realm_2021}, which recommended not to use this approach. They write, ``the fiducial absolute magnitude['s] [...] value depends on the method used to produce a light curve fit, which bands are included, the light curve age where it is defined, and the fiducial reference point chosen. Errors would arise from unintended mismatches between SN analyses and missing covariance data" \cite{di_valentino_realm_2021}. Lastly, Ref.~\cite{arendse_cosmic_2020} used strong gravitational lensing to break the degeneracy between $r_s$ and $H_0$ in the extrapolated BAO constraints; they found a small trend in the measured $r_d$ when using each lens (at different redshift) separately (see figure 5 there). While statistically insignificant ($1.6\sigma$) at the moment, this trend might signal residual systematics, either in the lenses themselves or in the procedure used to extrapolate the BAO measurements \cite{arendse_cosmic_2020}. Ref.~\cite{arendse_cosmic_2020} has noted that a recent ($z \approx 0.4$) change in dark energy might produce this behavior and re--absorb this trend.

\begin{figure}
    \centering
    \includegraphics{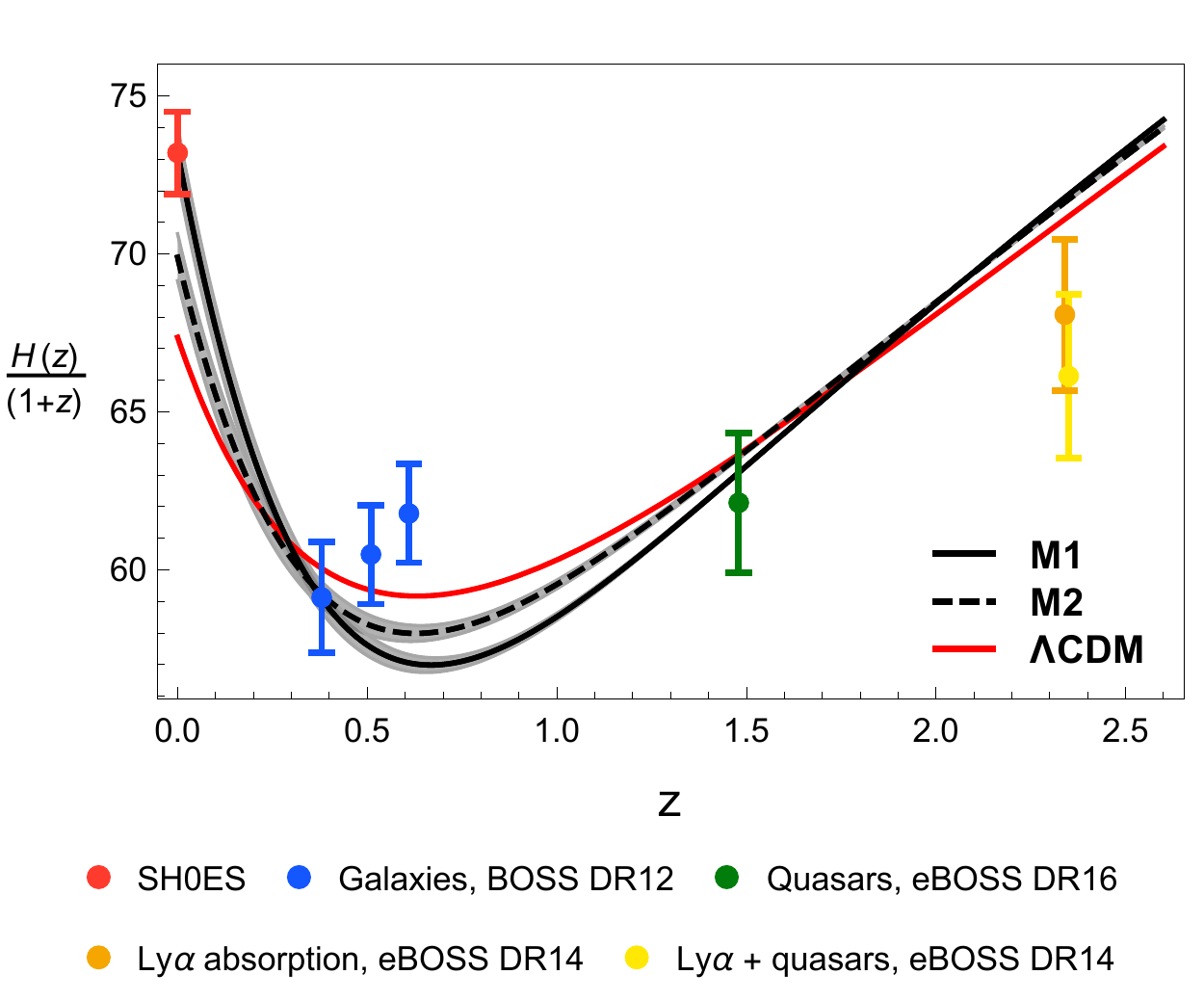}
    \caption{$H(z)/(1+z)$ in units of $\mathrm{km \: s^{-1} \: \! Mpc^{-1}}$ as a function of $z$. The theoretical predictions are shown: red curve -- \lcdm{} and black curves -- Lifshitz cosmology: unbroken -- M1 and dashed -- M2. The gray band around each black curve shows the $\pm 1 \sigma$ errors in $H(z)/(1+z)$ obtained by propagating the errors in $\Omega_{\infty} h^2$. Also shown are BAO results with $r_d = 147.09 \mathrm{\: Mpc}$ at several redshifts from: galaxy correlations in BOSS DR12 \cite{alam_clustering_2017}, quasar correlation in eBOSS DR16 \cite{hou_completed_2021}, the correlations of Ly$\alpha$ absorption in eBOSS DR14 \cite{agathe_baryon_2019}, and cross--correlation of Ly$\alpha$ absorption and quasars in eBOSS DR14 \cite{blomqvist_baryon_2019}. The SH0ES measurement at $z=0$ \cite{riess_cosmic_2021} is shown as well. While the overall fit of \lcdm{} to the BAO measurements seems to be somewhat better, the fit of the two Lifshitz cosmology realizations seems to be reasonably acceptable. The point at $z = 0.61$ already disagrees with \lcdm{}, but more severely so with M2 and even more with M1. The point at $z = 0.38$ agrees with Lifshitz cosmology (M1 and M2) slightly better, and so does the point at $z = 1.48$.}
    \label{BAOdata}
\end{figure}

Only a more thorough analysis of Lifshitz cosmology would be able to answer the question posed by extrapolating the BAO measurements; at the moment, we may take \emph{Planck}'s value for $r_d$ ($147.09 \mathrm{\: Mpc}$ \cite{planck_collaboration_planck_2020}) to see whether the Lifshitz cosmologies (M1 and M2) are consistent with the BAO measurements of $H(z)$ at $z \ge 0.38$. We take \lcdm{}'s value for the sound horizon at the end of the baryon--drag epoch to approximate the Lifshitz cosmology's value since $z_d$ shortly follows last--scattering ($z_* = 1,089.92$ \cite{planck_collaboration_planck_2020}). Figure~\ref{BAOdata} shows $H(z)/(1+z)$ at $z = 0.38, 0.51, 0.61, 1.48, 2.34,$ and $2.35$ from BAO measurements (with $r_d = 147.09 \mathrm{\: Mpc}$). In this figure, the data point at $z = 0.61$ already disagrees with \lcdm{}, but more severely so with M2 and even more with M1; the data point at $z = 0.38$ agrees with the Lifshitz cosmologies (M1 and M2) slightly better than with \lcdm{}, and so does the point at $z = 1.48$. By observing Fig.~\ref{BAOdata}, we conclude that while the overall fit of \lcdm{} to BAO measurements of $H(z)$ seems to be somewhat better, the fit of the Lifshitz cosmologies is acceptable.

\section{Early universe} \label{Implications_for_the_early_universe}

Up to this point, we have ignored the early universe and solved the theory assuming only late--universe modifications. Now, we shall turn our attention to this point. We will estimate the expected early--universe modifications due to Lifshitz cosmology to assess the validity of the assumption that led us to drop the radiation term. Specifically, we will verify that our Lifshitz cosmology's dynamics are consistent with negligible dark energy contribution at the early universe.

We return to the two coupled equations, \EqsRef{\ref{LC_self_consistent_dynamics}}, that describe the mutual interaction between the expanding universe and dark energy according to Lifshitz cosmology. This time, we do not drop the radiation term, and we need to find a new solution that includes this term. In addition, we can no longer assume that the sound horizon at last--scattering $r_*$ is unchanged. Therefore, if we wish to proceed in the spirit of Sec.~\ref{Approximate_Solution} and find a set of values for the theory's parameters by imposing a relationship between $\al$ and $\Omega_{\infty} h^2$; then, instead of demanding that $D_M$ is kept unchanged (\EqRef{\ref{fixing_the_integration_constant}}), we should demand that the CMB's angular acoustic scale $\theta_*$ is unchanged. That is, we should, in principle, demand
\begin{equation}
    \theta_* \equiv \frac{r_*}{D_M} = \theta_*^{(\Lambda \mathrm{CDM})}.
    \label{fixing_the_integration_constant_at_early_universe}
\end{equation}

As it turns out (see Appendix~\ref{Early_universe}), it is not straightforward to generalize our solution (\EqRef{\ref{First_order_H_after_equality}}) to accommodate $\Omega_r a^{-4}$ and then solve \EqRef{\ref{fixing_the_integration_constant_at_early_universe}}. Therefore, we will approximate by taking a detour: we take $\al$ and $\Omega_{\infty} h^2$ that we found for M1 and M2 when considering only late modifications (Table~\ref{parameters}), and then we plug (as the zeroth--order solution) $H^2 = H_0^2(\Omega_r a^{-4} + \Omega_m a^{-3} + \Omega_{\infty})$ into the equation for $H_0^2 \Dot{\Omega}_{LC}$ (the second equation in \EqsRef{\ref{LC_self_consistent_dynamics}}), and we integrate it (numerically, see Appendix~\ref{Early_universe}) with $\Omega_{\infty} h^2$ as the integration constant. This way, we get an approximation for the first--order (in $\al$) $\Omega_{LC} h^2$ that includes early--universe effects. Now, we can use the so--obtained $\Omega_{LC} h^2$ to calculate the relative contribution of the vacuum energy, $f_{de}$, throughout the entire cosmic evolution
\begin{equation}
    f_{de} \equiv \frac{\omega_{LC}}{\omega_r a^{-4} + \omega_m a^{-3} + \omega_{LC}}
    \label{f_de_definition}
\end{equation}
where $\omega_x \equiv \Omega_x h^2$ for $x = r,m,LC$.

\begin{figure}
    \centering
    \includegraphics{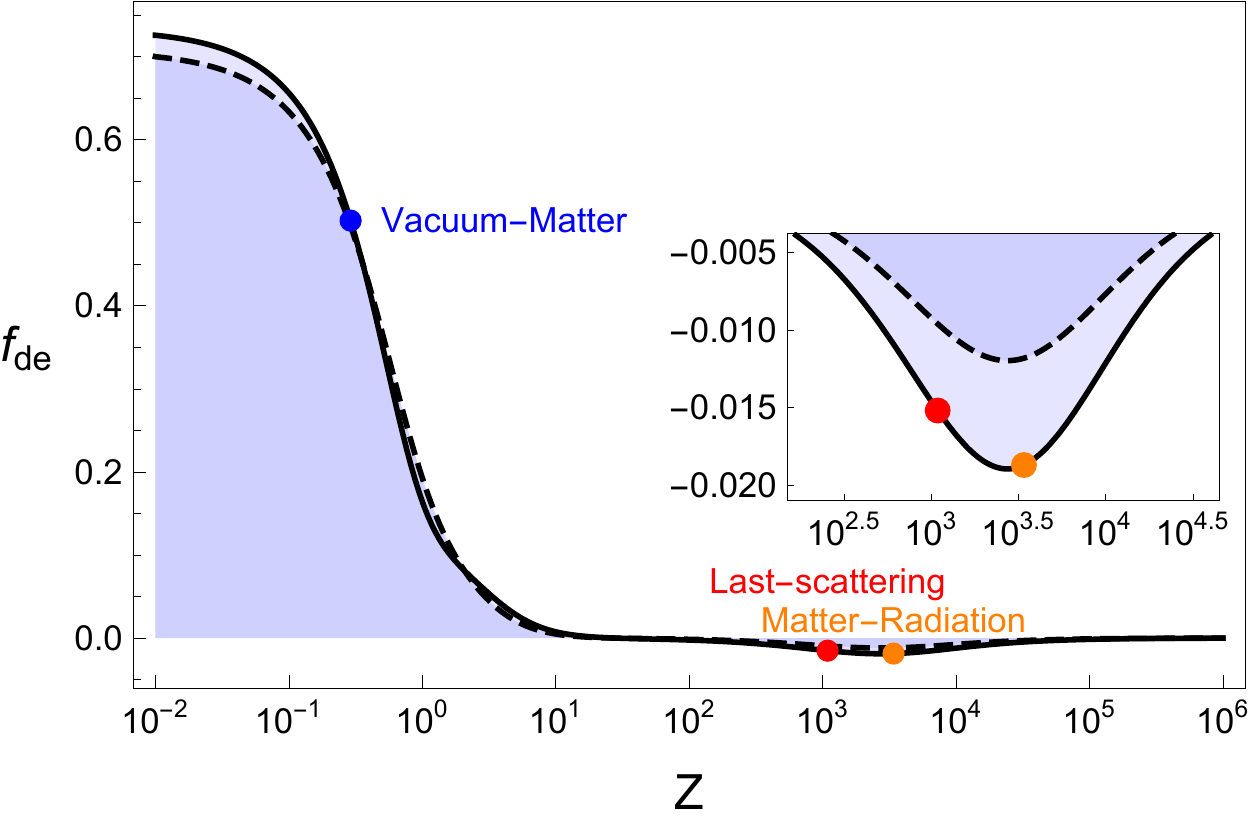}
    \caption{Relative dark energy contribution $f_{de}$ as a function of $z$, as calculated from the approximation of Lifshitz cosmology's dynamics including the early universe. The unbroken curve shows the case M1 and the dashed curve M2. The $\pm 1 \sigma$ errors in $f_{de}$ due to errors in $\Omega_{\infty} h^2$ are not shown here since they are thinner than the curve's width. Three special times are presented: matter--radiation equality $z_{eq} = 3,402$ \cite{planck_collaboration_planck_2020} (orange), last--scattering $z_* = 1,089.92$ \cite{planck_collaboration_planck_2020} (red), and vacuum--matter equality $z_{vm} \approx 0.29$ (calculated from $f_{de}(z=z_{vm}) = 0.5$) (blue). The early--universe evolution of $f_{de}$, according to Lifshitz cosmology, takes place roughly at the range $223 \le z \le 30,350$ for M1 or $388 \le z \le 17,950$ for M2 where $f_{de} \le -0.005$, and peaks around $z_{eq}$ with $f_{de} \approx -0.019$ at the peak for M1 or $\approx -0.012$ for M2 (inset). At late times, $f_{de}$ rises drastically and becomes at the present $f_{de}(z=0) \approx 0.733$ for M1 or $\approx 0.707$ for M2. Far in the future, it approaches $1$.}
    \label{RelContZ}
\end{figure}

The results are shown in Fig.~\ref{RelContZ}. As expected,\footnote{Recall that Lifshitz cosmology predicts dark energy evolution only in transition periods} the dark energy's dynamics kick in around the transition period from radiation to matter domination at $z_{eq} = 3,402$ \cite{planck_collaboration_planck_2020}. The early--universe evolution of $f_{de}$, according to Lifshitz cosmology, takes place roughly at the range (that includes both last--scattering and matter--radiation equality) $223 \le z \le 30,350$ for M1 or $388 \le z \le 17,950$ for M2 where $f_{de} \le -0.005$, and it peaks around $z_{eq}$ with $f_{de} \approx -0.019$ at the peak for M1 or $\approx -0.012$ for M2 (Fig.~\ref{RelContZ}'s inset). The dark energy's dynamics kick in again around the transition period from matter to vacuum domination at $z_{vm} \approx 0.29$ (calculated from $f_{de}(z=z_{vm}) = 0.5$). There, it rises drastically and becomes at the present $f_{de}(z=0) \approx 0.733$ for M1 or $\approx 0.707$ for M2. It finally approaches $1$, far in the future.

This solution, which considers the early features of dark energy, involves a further approximation in addition to first--order perturbation theory: the parameters are obtained from the late--universe dynamics. Therefore, we should be more careful with drawing cosmological conclusions, but only regard it as an indication of the validity of our assumption for the late--universe dynamics. If this approximate solution would have predicted more noticeable modifications around last--scattering, then our assumption would be in question; the fact that this is not the case makes our assumption sensible. As a side remark, we cautiously mention that Ref.~\cite{philcox_determining_2021} suggests that early modification to \lcdm{} should treat matter--radiation equality and last--scattering scales similarly to solve the Hubble tension; it seems that Lifshitz cosmology does just that.

\section{Discussion} \label{Discussion}

Viewing the universe as one giant ``dielectric medium'' with time--dependent refractive index and applying Lifshitz theory for calculating the vacuum energy inside the medium, one can find a physical description of dark energy \cite{leonhardt_lifshitz_2019}. This description is based on well--established and well--tested physics  \cite{landau_statistical_1980,leonhardt_case_2020} which makes it unique among all other models of dark energy. The theory comes with two free parameters, $\al$ and $\Omega_{\infty} h^2$ (replacing $\Omega_\Lambda h^2$ of \lcdm{}, such we have a total of seven parameters). We call this theory Lifshitz cosmology.

In this paper, we have investigated two realizations of Lifshitz cosmology; for each realization, we choose a value for the coupling parameter $\al$, and then, by demanding \EqRef{\ref{fixing_the_integration_constant}}, $\Omega_{\infty} h^2$ is fixed together with the predicted dynamics.

Our first considered realization (M1) is $\alpha_\Lambda^{M1} = \alpha_\Lambda^{TH} = (9\pi)^{-1}$ based on the assumption of only electromagnetic contribution to the vacuum energy with a sharp cut--off at exactly the Planck length. Amazingly, this naive theoretical prediction gives the SH0ES value for $H_0$ ($73.2 \; [\mathrm{km \: s^{-1} \: \! Mpc^{-1}}]$ at mean value). We may (and in some instances, we should \cite{benevento_can_2020,efstathiou_h0_2021}) view the Hubble tension as a tension between the SH0ES value for $M_B$ and the one obtained by calibrating the Pantheon data with \lcdm{} or using inverse distance ladders \cite{benevento_can_2020,efstathiou_h0_2021}. Table~\ref{RMSD_table} shows that M1 can considerably relieve the tension by $51\%$ at the best $M_B$ value; the relatively small difference in RMSD between the best $M_B$ and the SH0ES value suggests that M1 might even resolve this $M_B$ tension completely. M1 also seems to fit $E^{-1}$ measurements based on distance--ladder--independent SN Ia data (Fig.~\ref{Eofzdata}). On the other hand, while M1 appears to fit the shape of the binned SN Ia data (Fig.~\ref{muVSzFull}) at the lower redshift region ($z \sim 0.2$), it does not fit the shape at the higher redshift region (to the extent that this shape does not depend on the model). M1 also seems to fit BAO measurements of $H(z)$ only moderately (Fig.~\ref{BAOdata}).

Our second considered realization (M2) is $\alpha_\Lambda^{M2} = 0.0225$. This model seems to be the middle ground between M1 and \lcdm{}; it gives a nominal value of $H_0 = 69.9 \pm 0.7 \; [\mathrm{km \: s^{-1} \: \! Mpc^{-1}}]$, and it shrinks $\Delta M_B$ by only $\sim 19\%$. On the other hand, M2 fits all the $E^{-1}$ data points (Fig.~\ref{Eofzdata}), it perfectly fits the shape of the binned SN Ia data over the entire redshift range (Fig.~\ref{muVSzFull}), and its fit to BAO measurements of $H(z)$ is comparable to \lcdm{}'s and only slightly worse (Fig.~\ref{BAOdata}).

We have compared Lifshitz cosmology with astronomical data for the first time. There is certainly room for improvement and there are opportunities for further research. One problem with our analysis is that it treats $\al$ and $\Omega_{\infty} h^2$ as related via \EqRef{\ref{fixing_the_integration_constant}}, whereas they should be regarded as two independent parameters. By treating these parameters independently, a future analysis may yield better results, as we demonstrate in Appendix~\ref{Toy_model} with a toy model. One could also determine the best among the models by implementing the approximate solution we found here (\EqRef{\ref{First_order_H_after_equality}}) in numerical codes to perform a parameter fitting and likelihood analysis of the CMB together with other key cosmological data. In any case, our analysis already proves that Lifshitz cosmology deserves serious consideration.


\section*{Acknowledgments}
Our research was supported by the Israel Science Foundation, the Murray B. Koffler Professorial Chair and Aalto University.


\appendix

\section{The vacuum state} \label{The_vacuum_state}

\renewcommand{\theequation}{A\arabic{equation}}
\setcounter{equation}{0}

The equation governing the evolution of $\Omega_{LC}$, the second equation in \EqsRef{\ref{LC_self_consistent_dynamics}}, which we introduced in Sec.~\ref{Equations_of_motion} as
\begin{equation}
    H_0^2 \Dot{\Omega}_{LC} = 8\alpha_\Lambda H \partial_t^3 H^{-1},
    \label{dark_energy_dynamics_cosmic_time}
\end{equation}
was introduced in Ref.~\cite{leonhardt_lifshitz_2019} in a different form (See Eq.~(21) there):
\begin{equation}
    H_0^2 \Dot{\Omega}_{LC} = 8\alpha_\Lambda H \left( \partial_t^3 H^{-1} +H\partial_t^2 H^{-1} \right).
    \label{dark_energy_dynamics_conformal_time}
\end{equation}
The difference between these two forms of $\Dot{\Omega}_{LC}$ stems from different definitions of the cosmologically relevant vacuum state.

The vacuum state of a quantum field is defined as the state that gets annihilated by all of the annihilation operators. Each set of creation and annihilation operators is defined in a specific coordinate system \cite{fulling_nonuniqueness_1973}, and as a result, the \emph{definition} of the vacuum state also depends on that coordinate system \cite{fulling_nonuniqueness_1973}. Consider a state that gets annihilated by all the annihilation operators in one frame of reference and thus appears as a vacuum there; this same state might not be annihilated by all the annihilation operators in another frame and hence appear as an excited state there \cite{fulling_nonuniqueness_1973}. This fact sometimes goes unappreciated or misunderstood, but it is known for a long time now \cite{fulling_nonuniqueness_1973}. The best--known example is the Unruh--Fulling--Davies effect \cite{fulling_nonuniqueness_1973,davies_scalar_1975,unruh_notes_1976}, where the quantum vacuum defined with respect to creation and annihilation operators in an inertial frame in Minkowski space appears in an accelerated frame as thermal radiation --- not as a vacuum.

This frame dependence means that a state defined as a vacuum in one frame is not necessarily the same state defined as a vacuum in another frame; these two states could be different from one another, i.e., two different physical settings. Let us emphasize that the general coordinate invariance does not break; instead, the ``vacuumness" of a quantum state is a frame--dependent quality. One may understand this frame--dependence with an analogy to a point--mass at rest.

The rest frame of a point--mass is one unique frame (up to translations), in which the point--mass appears to sit at rest; in other frames, the same point--mass appears to move. Of course, the physics describing the point--mass is independent of the frame in which we choose to observe it. Nonetheless, two point--masses which appear at rest in two different frames (not related by translations), do not represent the same physical system but two different systems. In this analogy, the ``vacuumness" of a quantum field is akin to the ``restness" of a point--mass. To conclude this idea, two quantum states defined as a vacuum in two different frames are two different physical states; they are not one and the same state observed in two different coordinates.

After clarifying this point, one question is raised when considering Lifshitz cosmology: Which is the relevant frame for defining the universe's vacuum state?

Ref.~\cite{leonhardt_lifshitz_2019} defined the vacuum state with respect to conformal time $\tau$,
\begin{equation}
    \tau = \int \frac{dt}{a(t)}.
    \label{conformal_time}
\end{equation}
In conformal time, the FLRW metric becomes conformally flat. As Maxwell's equations are conformally invariant \cite{birrell_quantum_1982}, the electromagnetic field and its fluctuations perceive the conformally flat expanding universe as flat Minkowski space with constant Hamiltonian and hence, an exact ground state. For this reason, Ref.~\cite{leonhardt_lifshitz_2019} thought to define the cosmological vacuum as a vacuum state with respect to conformal time, which leads to \EqRef{\ref{dark_energy_dynamics_conformal_time}}. We performed the same analysis as in Sec.~\ref{Approximate_Solution}, with \EqRef{\ref{dark_energy_dynamics_conformal_time}} replacing \EqRef{\ref{dark_energy_dynamics_cosmic_time}} in \EqsRef{\ref{LC_self_consistent_dynamics}}. We found that Lifshitz cosmology with conformal vacuum state leads to $H_0 < 67 \, [\mathrm{km \: s^{-1} \: \! Mpc^{-1}}]$ for any $\al$ in the range of interest. We thus conclude that the original version of Lifshitz cosmology, with a conformal vacuum state, is ruled out by the Hubble tension (that demands a higher $H_0$).

In this paper, with hindsight, we have tried to define the cosmological vacuum as a vacuum state with respect to cosmological time $t$. This definition of a vacuum state seems more natural as it means that the cosmological vacuum is co--moving with the expanding universe alongside anything else, matter and radiation alike. By so defining the vacuum state, we obtain \EqRef{\ref{dark_energy_dynamics_cosmic_time}}. Our current work shows that this definition seems able to resolve (or considerably mitigate) the Hubble tension. We thus conclude that the cosmologically relevant vacuum state should be defined as a vacuum with respect to cosmological time $t$.


\section{Calculations} \label{calculations}

\renewcommand{\theequation}{B\arabic{equation}}
\setcounter{equation}{0}

In this appendix, we will detail the calculations that were briefly described in sections \ref{Approximate_Solution} and \ref{Implications_for_the_early_universe}.

\subsection{Late universe} \label{Late_universe}

To obtain the late--universe evolution, we start from \EqsRef{\ref{LC_self_consistent_dynamics}}, which we write again here:
\begin{equation}
    \Bigg\{
    \begin{aligned}
        H^2(a) \;\! \ &= H_0^2(\Omega_r a^{-4} + \Omega_m a^{-3} + \Omega_{LC}), \\
        H_0^2 \Dot{\Omega}_{LC} &= 8\alpha_\Lambda H \partial_t^3 H^{-1}.
    \end{aligned}
    \label{LC_self_consistent_dynamics_appendix}
\end{equation}

These two coupled equations describe the mutual interaction between the cosmic expansion and the evolution of dark energy. By solving these equations, we can find the evolution of the background universe (homogeneous and isotropic) according to Lifshitz cosmology. The problem is that these equations are not easy to solve. As mentioned in Sec.~\ref{Approximate_Solution}, even obtaining a reliable numerical solution is considerably hard for the following two reasons. First, the equation for the dark energy's dynamics, the second equation in \EqsRef{\ref{LC_self_consistent_dynamics_appendix}}, depends on high derivatives of the scale--factor $a$ (up to fourth--order derivative); this is a problem because, in differential equation solvers, the highest derivatives take the lead, whereas in reality, for most of the period of interest, the dynamics of $\Omega_{LC}$ are a mere correction to the dynamics of the universe. Second, \EqsRef{\ref{LC_self_consistent_dynamics_appendix}} constitute what is known as ``stiff equations," causing havoc with step size and accuracy. Therefore, we will approximate and solve perturbatively, where $\al$ will be our small parameter. As also discussed in Sec.~\ref{Approximate_Solution}, our first simplification will be dropping the radiation term $H_0^2\Omega_r a^{-4}$.

Next, the combination $H_0^2\Omega_m$ that appears in the first equation of \EqsRef{\ref{LC_self_consistent_dynamics_appendix}} is proportional to the present--day physical density of matter $\rho_{0,m}$; as this density is a physical entity, $H_0^2\Omega_m$ should be a model--independent combination. Indeed, $H_0^2\Omega_m$ can be determined model--independently by the relative heights of the CMB acoustic peaks \cite{planck_collaboration_planck_2014}. Therefore, we may replace $\omega_m \equiv \Omega_m h^2$ (where $h \equiv H_0/100 [\mathrm{km \: s^{-1} \: \! Mpc^{-1}}]$) in \EqsRef{\ref{LC_self_consistent_dynamics_appendix}} by its \lcdm{}'s equivalent
\begin{align}
    \nonumber \omega_m^P \equiv [\Omega_m h^2]^P &= 0.1430 \pm 0.0011 \\
    &\qquad \quad \text{(\emph{Planck}, TT,TE,EE+lowE+lensing)},
    \label{omega_m_planck}
\end{align}
where $[\Omega_m h^2]^P$ is obtained by \emph{Planck}'s TT,TE,EE+lowE+lensing \lcdm{} analysis \cite{planck_collaboration_planck_2020} (the `\textsc{p}' superscript denotes that we use \emph{Planck}'s \lcdm{} value). In the following, we will also use \emph{Planck}'s value \cite{planck_collaboration_planck_2020} of
\begin{align}
    \nonumber \omega_\Lambda^P \equiv [\Omega_\Lambda h^2]^P &= 0.3107 \pm 0.0082 \\
    &\qquad \quad \text{(\emph{Planck}, TT,TE,EE+lowE+lensing)}.
    \label{omega_lambda_planck}
\end{align}

Now, for mathematical convenience, we re--scale and re--define the variables
\begin{equation}
    \nu \equiv \ln{[(\omega_\Lambda^P/\omega_m^P)^{1/3} a]}, \quad \xi \equiv \sqrt{\omega_\Lambda^P} \, t, \quad \eta \equiv \frac{\omega_{LC}}{\omega_\Lambda^P} \, ,
    \label{rescaling_of_a_and_t_and_Omega_LC}
\end{equation}
where $\omega_{LC} \equiv \Omega_{LC} h^2$. We also regard $\xi$ (time) as a function of $\nu$ (scale--factor) and define $\Theta$ as the derivative of $\xi$ with respect to $\nu$. It is easy to show that $\Theta = \sqrt{\omega_\Lambda^P} \, H^{-1}$ (expressed in terms of $a$ and $t$):
\begin{equation}
    \Theta \equiv \frac{d\xi}{d\nu} = \Big(\frac{d\nu}{d\xi}\Big)^{-1} = \Big(\frac{da}{d\xi}\frac{d\nu}{da}\Big)^{-1} = \sqrt{\omega_\Lambda^P}\Big(\frac{da}{dt}\frac{1}{a}\Big)^{-1} = \sqrt{\omega_\Lambda^P} \, H^{-1}.
    \label{theta_def}
\end{equation}
Then, \EqsRef{\ref{LC_self_consistent_dynamics_appendix}} become (dropping the radiation term and replacing $\omega_m$ with $\omega_m^P$)
\begin{align}
    \label{theta_dynamics} \theta &= (\mathrm{e}^{-3\nu}+\eta)^{-1/2}, \\
    \label{eta_dynamics} \eta' &= 8\al \frac{1}{\theta^2} \partial_{\nu}\bigg(\partial_{\nu}\frac{\theta'}{\theta}-\frac{1}{2}\frac{\theta'^2}{\theta^2}\bigg)
\end{align}
where $\theta \equiv 100 [\mathrm{km \: s^{-1} \: \! Mpc^{-1}}] \Theta$ and the primes indicate differentiation with respect to $\nu$.

At this point, we notice that given a solution $\{\theta_a(\nu), \ \eta_a(\nu)\}$ of \EqsRef{\ref{theta_dynamics}} and (\ref{eta_dynamics}), it is easy to show that $\{\theta_b(\nu), \ \eta_b(\nu)\}$ is also a solution, where
    \begin{align}
        \label{scaling_and_shifting_symmetry} \theta_b(\nu) &= \theta_a(\nu-\delta)\mathrm{e}^{(3/2)\delta}, \\
        \label{scaling_and_shifting_symmetry_of_eta} \eta_b(\nu) &= \eta_a(\nu-\delta) \mathrm{e}^{-3\delta}
    \end{align}
for some constant $\delta$. One immediate result is that from any solution $\{\theta_a(\nu), \ \eta_a(\nu)\}$ with asymptotic behavior $\eta_a \rightarrow 1$ for $\nu \rightarrow +\infty$, we may construct a solution $\{\theta_b(\nu), \ \eta_b(\nu)\}$ with any other constant asymptotic behavior by choosing an appropriate value for $\delta$, as may be seen from
\begin{equation}
    \lim_{\nu\rightarrow+\infty}\eta_b(\nu) = \lim_{\nu\rightarrow+\infty}\eta_a(\nu-\delta) \mathrm{e}^{-3\delta} = \mathrm{e}^{-3\delta}.
    \label{asymptotic_behavior}
\end{equation}
This relation translates $\delta$ to the integration constant in \EqRef{\ref{eta_dynamics}} when integrating $\eta$ from far in the future ($\nu \rightarrow +\infty$) to some other value $\nu$; thus, we can solve the equations with a convenient integration constant and later find the physically relevant one by adjusting $\delta$. This symmetry also translates $\delta$ to a value of $\omega_{LC}$ far in the future
\begin{equation}
    \omega_{\infty} = \omega_\Lambda^P \mathrm{e}^{-3\delta},
    \label{asymptotic_behavior_of_Omega_as_function_of_delta}
\end{equation}
where $\omega_{\infty} \equiv \displaystyle \lim_{\nu\rightarrow+\infty}\omega_{LC}(\nu)$. Thus, any solution $\theta(\nu;\al,\delta)$ is characterized by two parameters $\al$ and $\delta$.

Now, we proceed in two steps. The first one is to find $\eta$ up to first--order in $\al$ with $\eta_{\infty} \equiv \displaystyle \lim_{\nu\rightarrow+\infty}\eta = 1$ and then use it to find $\theta$. The second step is finding the value of $\delta$ by demanding that the angular diameter distance to the surface of last--scattering, $D_M$, is unchanged by Lifshitz cosmology (\EqRef{\ref{fixing_the_integration_constant}}).

For the first step, we find $\theta^{(0)}$ for $\al=0$ (constant $\eta$) and $\eta^{(0)} = \eta_{\infty} = 1$ (equivalent to $\delta = 0$) from \EqRef{\ref{theta_dynamics}}:
\begin{equation}
    \theta^{(0)} = (\mathrm{e}^{-3\nu}+1)^{-1/2}.
    \label{zeroth_order_theta}
\end{equation}
Then, we plug $\theta^{(0)}$ into \EqRef{\ref{eta_dynamics}} and integrate it with $\eta_{\infty}^{(1)} = 1$ to find $\eta^{(1)}$:
\begin{equation}
    \eta^{(1)} = 1+18\al\bigg[\ln{(\mathrm{e}^{-3\nu}+1)}-\frac{3}{\mathrm{e}^{3\nu}+1}\bigg].
    \label{first_order_eta}
\end{equation}
Finally, we plug $\eta^{(1)}(\nu)$ in \EqRef{\ref{theta_dynamics}} to find $\theta^{(1)}(\nu)$ and then use $\theta^{(1)}(\nu)$ in \EqRef{\ref{scaling_and_shifting_symmetry}} to find a general and adjustable (first--order) solution:
\begin{equation}
    \begin{aligned}
        &\theta(\nu;\al,\delta) = \theta^{(1)}(\nu-\delta)\mathrm{e}^{(3/2)\delta} \\
        &\qquad \, = \frac{\mathrm{e}^{(3/2)\delta}}{\sqrt{\mathrm{e}^{-3(\nu-\delta)}+1+18\al\bigg[\ln{(\mathrm{e}^{-3(\nu-\delta)}+1)}-\frac{3}{\mathrm{e}^{3(\nu-\delta)}+1}\bigg]}}.
    \end{aligned}
    \label{first_order_theta}
\end{equation}
We do not go beyond the first--order since, as it turns out, the expansion in $\al$ is a divergent series. By undoing the re--scalings and re--definitions of \EqsRef{\ref{rescaling_of_a_and_t_and_Omega_LC}}, taking $\theta = 100 [\mathrm{km \: s^{-1} \: \! Mpc^{-1}}] \sqrt{\omega_\Lambda^P} \, H^{-1}$, and reinstating the original $\omega_m$, we get \EqRef{\ref{First_order_H_after_equality}}.

For the second step, let us express $D_M$ with our terminology here:
\begin{equation}
    D_M = \frac{c \, (\omega^P_{\Lambda})^{-\frac{1}{6}} (\omega^P_m)^{-\frac{1}{3}}}{100 [\mathrm{km \: s^{-1} \: \! Mpc^{-1}}]} \int^{\nu_0}_{\nu_*} \theta(\nu';\al,\delta) \mathrm{e}^{-\nu'} d\nu',
    \label{D_M_in_rescaled_terminology}
\end{equation}
where $\ \nu_0 = \frac{1}{3} \ln{[\omega_\Lambda^P/\omega_m^P]}$, $\ \nu_* = \nu_0  - \ln{(1+z_*)}$. Hence, the demand $D_M \stackrel{!}{=} D_M^{(\Lambda \mathrm{CDM})}$ becomes
\begin{equation}
    \int^{\nu_0}_{\nu_*} \theta(\nu';\al,\delta) \mathrm{e}^{-\nu'} d\nu' \stackrel{!}{=} \int^{\nu_0}_{\nu_*} \theta(\nu';0,0) \mathrm{e}^{-\nu'} d\nu',
    \label{equal_D_M_in_rescaled_terminology}
\end{equation}
as $\al = 0$ and $\delta = 0$ gives the \lcdm{} solution. Lastly, since for times before last--scattering Lifshitz cosmology and \lcdm{} (approximately) coincide, we may take, for numerical simplicity, the lower limit of the integration on both sides of \EqRef{\ref{equal_D_M_in_rescaled_terminology}} to minus infinity:
\begin{equation}
    \begin{aligned}
        \int^{\nu_0}_{-\infty} \theta(\nu';\al,\delta) \mathrm{e}^{-\nu'} d\nu' &\stackrel{!}{=} \int^{\nu_0}_{-\infty} (\mathrm{e}^{-3\nu'}+1)^{-1/2} \mathrm{e}^{-\nu'} d\nu' \\
        &= 2 \bigg(\frac{\omega^P_{\Lambda}}{\omega^P_m}\bigg)^{1/6}{_2F_1}\bigg(\frac{1}{6},\frac{1}{2},\frac{7}{6},-\frac{\omega^P_{\Lambda}}{\omega^P_m}\bigg)
    \end{aligned}
    \label{fitting_condition_for_delta}
\end{equation}
where $\theta(\nu;\al,\delta)$ is given by \EqRef{\ref{first_order_theta}} and ${_2F_1}$ is Gauss' hypergeometric function. We can numerically solve \EqRef{\ref{fitting_condition_for_delta}} for a given $0<\al\ll1$ to obtain the corresponding value of $\delta$. (For a given $0<\al\ll1$, $\delta$ is a monotonic function of $D_M$.) Thus, we get the relationship presented in Fig.~\ref{OmegaInfVSalpha} (where we translated $\delta$ to $\omega_{\infty}$).

To estimate the errors in $\delta$ (or equivalently, $\omega_{\infty}$) that we obtain with the procedure above, we estimate the errors in $D_M^{(\Lambda \mathrm{CDM})} = r_*^P/\theta_*^P$ from \emph{Planck}'s measurement as
\begin{equation}
    \sigma_{D_M} = \sqrt{\left( \frac{\partial D_M}{\partial r_*} \bigg|_{r_*^P} \cdot \sigma_{r_*}^P \right)^2 +\left( \frac{\partial D_M}{\partial \theta_*} \bigg|_{\theta_*^P} \cdot \sigma_{\theta_*}^P \right)^2}.
\end{equation}
And according to \EqRef{\ref{D_M_in_rescaled_terminology}} we get
\begin{equation}
    s_{D_M} = \left(\frac{c \, (\omega^P_{\Lambda})^{-\frac{1}{6}} (\omega^P_m)^{-\frac{1}{3}}}{100 [\mathrm{km \: s^{-1} \: \! Mpc^{-1}}]}\right)^{-1} \sigma_{D_M} \approx 0.00363 \, ,
\end{equation}
where we use \emph{Planck}'s values \cite{planck_collaboration_planck_2020}. Then, we calculate the errors in $\delta$ as $\sigma_\delta^\pm = |\delta - \delta_\pm|$, where $\delta_+$ is calculated by adding $s_{D_M}$ to the right--hand side of \EqRef{\ref{fitting_condition_for_delta}} and solving for $\delta$, and $\delta_-$ is calculated by subtracting $s_{D_M}$ from the right--hand side. (This estimation procedure makes sense as $\delta$ is a monotonic function of $D_M$.) These errors in $\delta$ (or $\omega_{\infty}$) are estimates solely based on \emph{Planck}'s errors in $D_M$ and are probably underestimated.


\subsection{Early universe} \label{Early_universe}

To see how Lifshitz cosmology modifies the early dynamics, we must go back to \EqsRef{\ref{LC_self_consistent_dynamics_appendix}} and begin all over again without dropping the radiation term. This term complicates matters. For example, the scaling and shifting symmetry of \EqsRef{\ref{scaling_and_shifting_symmetry}} and (\ref{scaling_and_shifting_symmetry_of_eta}) does not hold anymore, and adjusting the physical value of the integration constant becomes difficult. Another issue is that we can no longer assume that the sound horizon at last--scattering $r_*$ is unchanged. Therefore, to find the integration constant, keeping the spirit of our analysis of the late universe, we should demand that the CMB's angular acoustic scale $\theta_* \equiv r_*/D_M$ is unchanged: $\theta_* \stackrel{!}{=} \theta_*^{(\Lambda \mathrm{CDM})}$ (instead of demanding $D_M \stackrel{!}{=} D_M^{(\Lambda \mathrm{CDM})}$ as before).

To go around these issues, we will approximate by building upon our late--universe result. This time, we remain with $t$ and $\omega_{LC}$ without re--scaling, and instead of $\nu$, we take
\begin{equation}
    \Tilde{\nu} \equiv \ln{a},
    \label{early_rescaling_of_a_and_t}
\end{equation}
and define $\Tilde{\Theta} \equiv dt/d\Tilde{\nu} = H^{-1}$ and $\Tilde{\theta} \equiv 100 [\mathrm{km \: s^{-1} \: \! Mpc^{-1}}] \Tilde{\Theta}$, then we get
\begin{align}
    \label{Tilde_theta_dynamics} \Tilde{\theta} &= (\omega_r \mathrm{e}^{-4\Tilde{\nu}} + \omega_m^P \mathrm{e}^{-3\Tilde{\nu}} + \omega_{LC})^{-1/2}, \\
    \label{Tilde_omega_dynamics} \omega'_{LC} &= 8\al \frac{1}{\Tilde{\theta}^2} \partial_{\Tilde{\nu}}\bigg(\partial_{\Tilde{\nu}}\frac{\Tilde{\theta}'}{\Tilde{\theta}}-\frac{1}{2}\frac{\Tilde{\theta}'^2}{\Tilde{\theta}^2}\bigg),
\end{align}
where now primes indicate differentiation with respect to $\Tilde{\nu}$, and $\omega_r$ can be determined (model--independently) from the Stefan--Boltzmann law of black--body radiation with the present--day average CMB temperature ($T_0 = 2.7255 \mathrm{K}$ \cite{fixsen_temperature_2009}):
\begin{equation}
    \omega_r = \frac{8\pi G}{(100 [\frac{\mathrm{km \: s^{-1}}}{\mathrm{Mpc}}])^2}\bigg[1+3\frac{7}{8}\bigg(\frac{4}{11}\bigg)^{4/3}\bigg]\frac{\pi^2(k_BT_0)^4}{15c^5\hbar^3} = 4.15\cdot 10^{-5} \, .
    \label{omega_r_stefan_boltzmann}
\end{equation}

At this point, we use our late--universe result. For the zeroth--order solution ($\al = 0$), we take $\Tilde{\theta}^{(0)} = (\omega_r \mathrm{e}^{-4\Tilde{\nu}} + \omega_m^P \mathrm{e}^{-3\Tilde{\nu}} + \omega_{\infty})^{-1/2}$ with $\omega_{\infty} = \omega_\Lambda^P \mathrm{e}^{-3\delta}$ (\EqRef{\ref{asymptotic_behavior_of_Omega_as_function_of_delta}}), where we use the value of $\delta$ we found in the late--universe solution. Then, we find $\omega_{LC}$ up to first--order in $\al$ by integrating \EqRef{\ref{Tilde_omega_dynamics}} with $\Tilde{\theta}^{(0)}$ using $\omega_{\infty}$ as the integration constant:
\begin{equation}
    \omega_{LC} = 8\al \int^{\ln{a}}_{\infty} \frac{1}{(\Tilde{\theta}^{(0)})^2} \partial_{\Tilde{\nu}}\bigg(\partial_{\Tilde{\nu}}\frac{(\Tilde{\theta}^{(0)})'}{\Tilde{\theta}^{(0)}}-\frac{1}{2}\frac{(\Tilde{\theta}^{(0)})'^2}{(\Tilde{\theta}^{(0)})^2}\bigg) d\Tilde{\nu} + \omega_{\infty}.
    \label{early_omega}
\end{equation}

Then, we may use $\omega_{LC}$ to calculate the relative contribution of the vacuum energy, $f_{de}$, throughout the entire evolution of the cosmos:
\begin{equation}
    f_{de} \equiv \frac{\omega_{LC}}{\omega_r a^{-4} + \omega_m^P a^{-3} + \omega_{LC}}.
    \label{f_de_definition_appendix}
\end{equation}

The results are shown in Fig.~\ref{RelContZ}.



\section{Toy model} \label{Toy_model}

\renewcommand{\theequation}{C\arabic{equation}}
\setcounter{equation}{0}

\begin{figure}[h]
    \centering
    \includegraphics[width=0.8\textwidth]{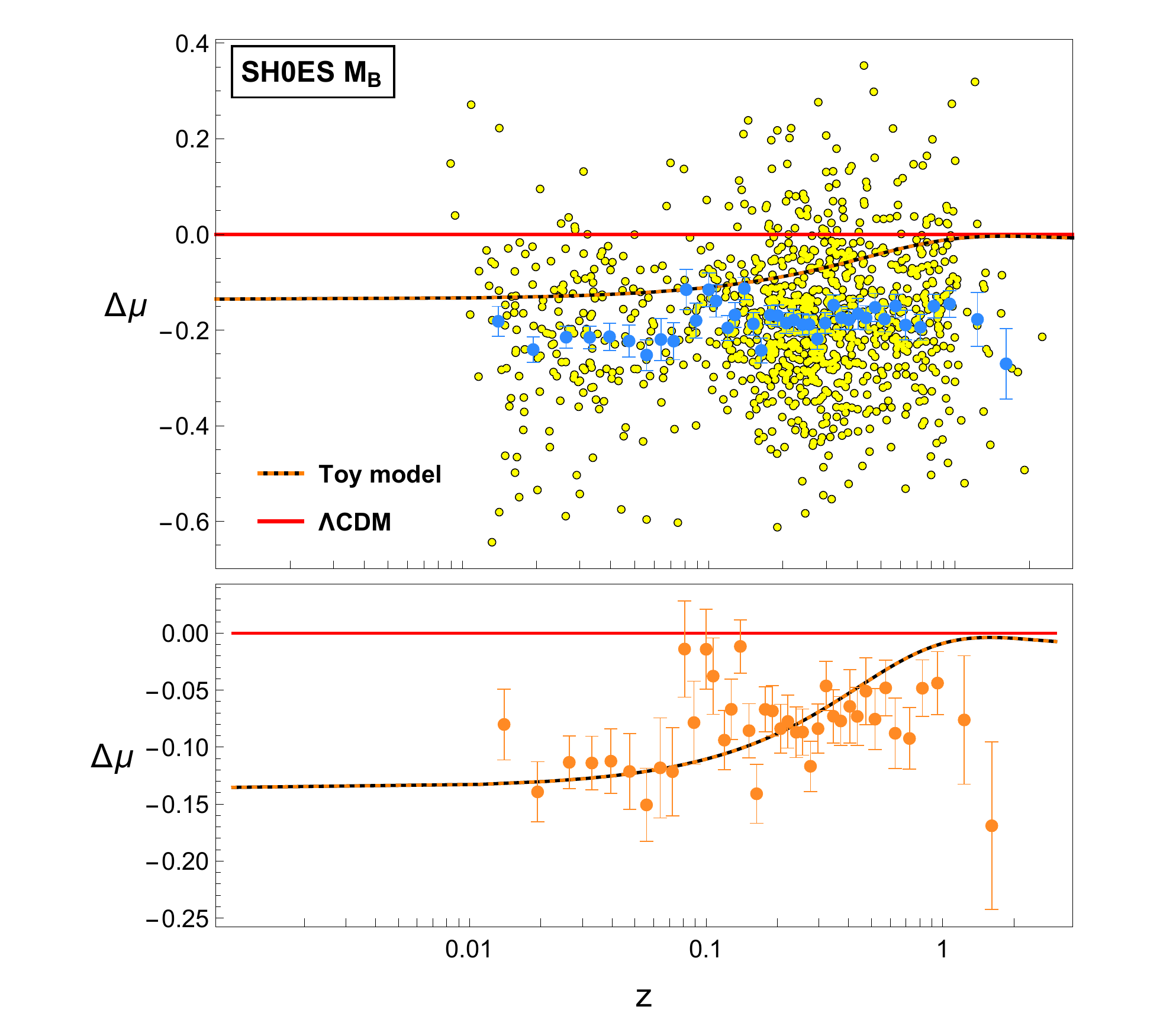}
    \caption{Same as Fig.~\ref{muVSzFull}; showing the prediction by the toy model ($\al = 0.025$ and $\Omega_{\infty} h^2 = 0.4625$), see Appendix~\ref{Toy_model}. The orange points at the bottom panel show the binned Pantheon data with $M_B = -19.346$, the best $M_B$ in terms of RMSD (\EqRef{\ref{RMSD_equation}}) for the Toy model.}
    \label{muVSzFullToyv6}
\end{figure}

\begin{figure}
    \centering
   \includegraphics{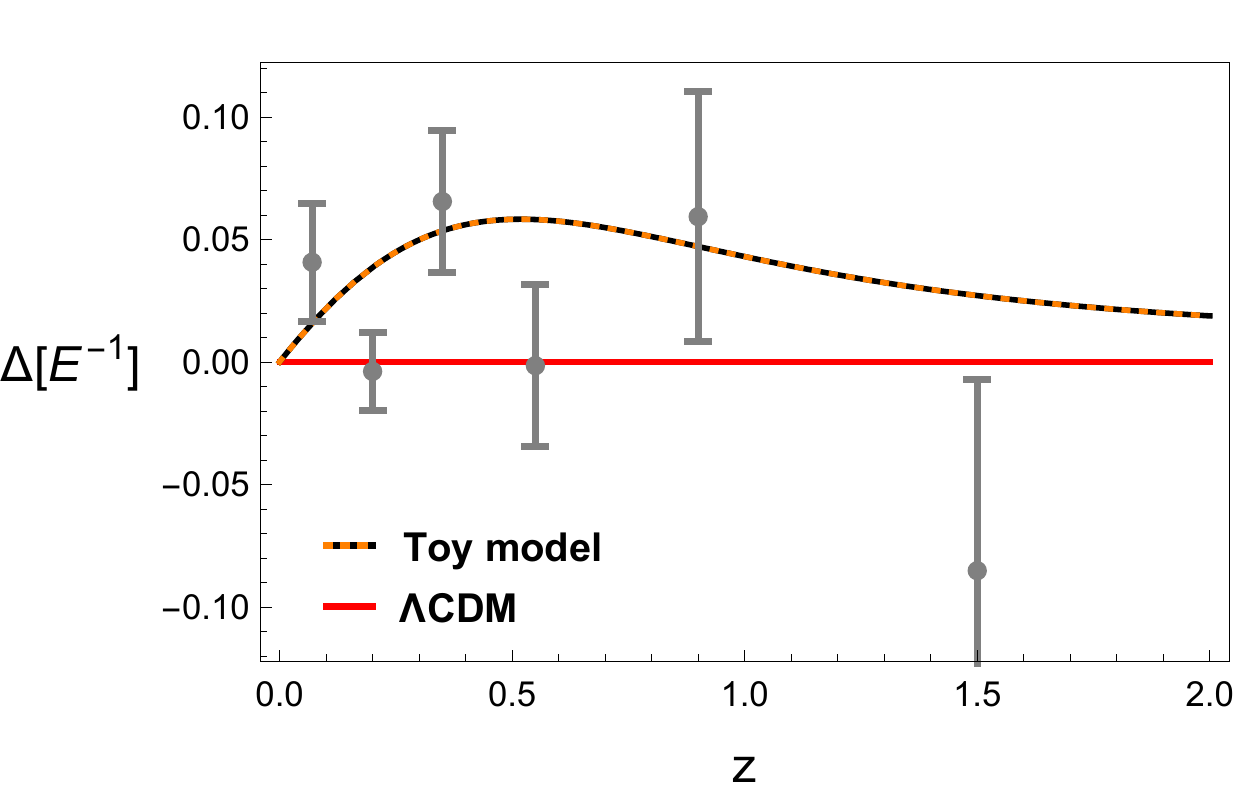}
    \caption{Same as Fig.~\ref{Eofzdata}; showing the prediction by the toy model ($\al = 0.025$ and $\Omega_{\infty} h^2 = 0.4625$), see Appendix~\ref{Toy_model}.}
    \label{EofzdataToyv6}
\end{figure}

\begin{figure}
    \centering
    \includegraphics{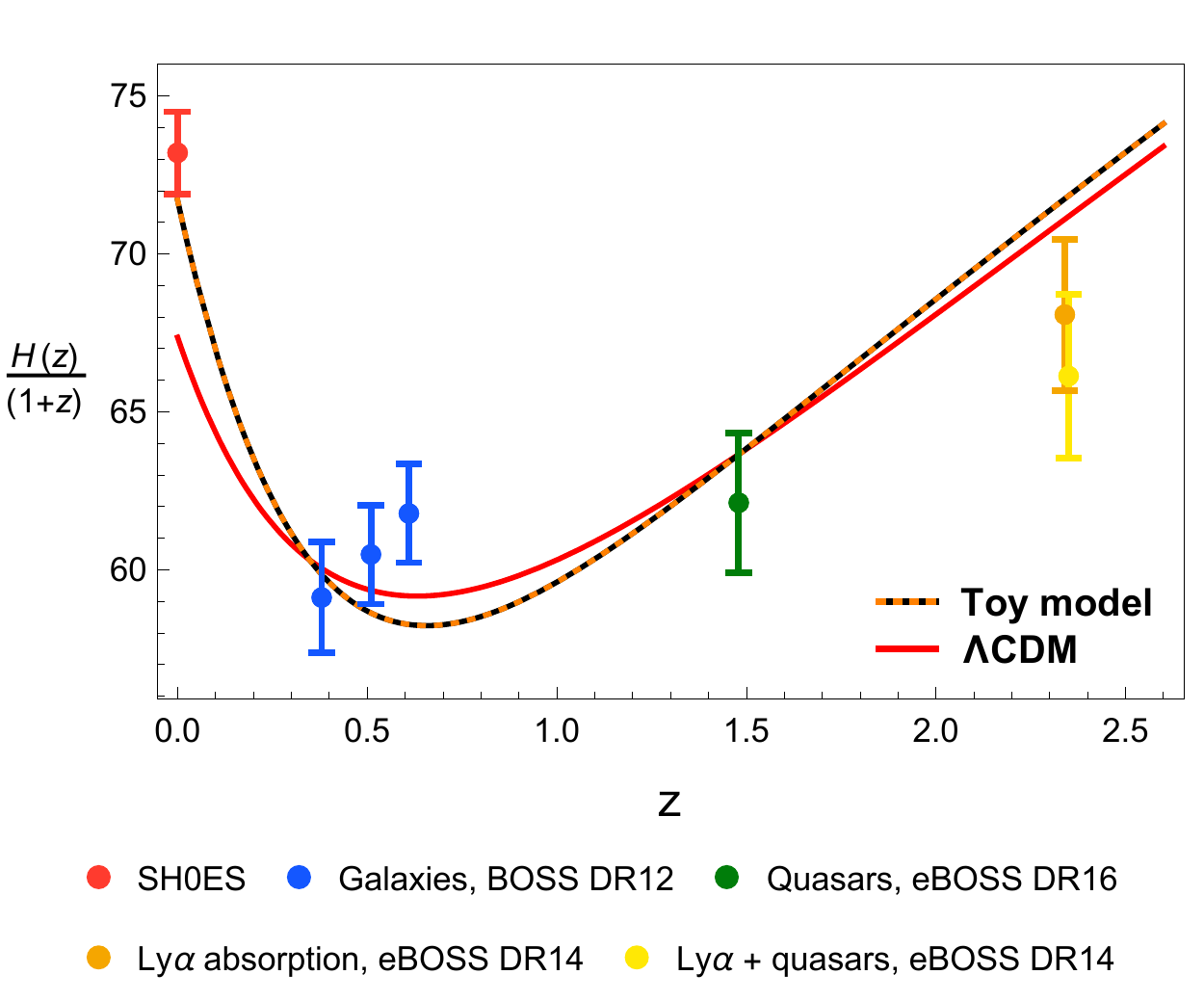}
    \caption{Same as Fig.~\ref{BAOdata}; showing the prediction by the toy model ($\al = 0.025$ and $\Omega_{\infty} h^2 = 0.4625$), see Appendix~\ref{Toy_model}.}
    \label{BAOdataToyv6}
\end{figure}

In this paper, we have introduced and analyzed for the first time Lifshitz cosmology with the specific goals of solving for the theory's dynamics and testing the viability of this theory. Therefore, we have avoided the more complex statistical analysis that must be eventually done to draw firm conclusions; this analysis is left open for future research. As we mentioned in the discussion section (Sec.~\ref{Discussion}), our analysis treats $\al$ and $\Omega_{\infty} h^2$ as two mutually dependent parameters related via \EqRef{\ref{fixing_the_integration_constant}}, whereas in reality, they are two independent parameters. By treating these parameters independently, a future analysis may yield better results. To demonstrate this idea, we arbitrarily choose the combination $\al = 0.025$ and $\Omega_{\infty} h^2 = 0.4625$, which produces $D_M$ that is (statistically insignificant) $1.7\sigma$ away from its \emph{Planck} value; we call this realization `toy model.' This toy model gives a nominal value of $H_0 = 71.7 \; [\mathrm{km \: s^{-1} \: \! Mpc^{-1}}]$, and it shrinks $\Delta M_B$ by about $43\%$ ($M_B^* = -19.346$) with $\mathrm{RMSD} = 0.1466$, and thus it performs better than our M2. In addition, the toy model seems to fit the shape of the binned SN Ia data, the distance--ladder--independent $E^{-1}$ measurements from SN Ia, and the BAO measurements of $H(z)$ better than M1 (and comparably or better than M2), see Figs.~\ref{muVSzFullToyv6}, \ref{EofzdataToyv6}, and \ref{BAOdataToyv6}. Overall, the toy model seems to work better than M1 and M2 and might be closer to the truth. Rigorous statistical analysis will probably produce an even better realization.


\vspace{2.5cm}

\newpage

\printbibliography[title={References}]


\end{document}